\documentclass[preprint,3p,times]{elsarticle}

\usepackage{amsmath}
\usepackage{amssymb}
\usepackage{graphicx}
\usepackage{booktabs}
\usepackage{array}
\usepackage{caption}
\usepackage{subcaption}
\usepackage{enumitem}
\usepackage{placeins}
\usepackage{float}
\usepackage{longtable}
\usepackage{pdflscape}
\usepackage{xcolor}
\usepackage[hidelinks]{hyperref}

\graphicspath{{./figures/}{./ch5_results/}}
\journal{Applied Energy}

\newif\ifshowtodos
\showtodostrue          % <-- comment out this line before submission

\begin{document}

\begin{frontmatter}

\title{Tail reliability in probabilistic electricity price forecasts: comparing
model combination and conformal prediction in the GB day-ahead market}

\author[ucl]{Anirudh Agrawal}
\author[ucl]{Aidan O'Sullivan\corref{cor1}}
\ead{aidan.osullivan@ucl.ac.uk}
\cortext[cor1]{Corresponding author.}

\affiliation[ucl]{organization={UCL Energy Institute, The Bartlett Faculty of the
Built Environment, University College London},
  addressline={14 Upper Woburn Place},
  city={London},
  postcode={WC1H 0NN},
  country={United Kingdom}}

\begin{abstract}
Probabilistic forecasts of day-ahead electricity prices are increasingly used to
manage price risk, yet competing models are usually ranked on aggregate scores
that average over the price regimes in which uncertainty matters most. This
study evaluates probabilistic forecasts of hourly EPEX SPOT GB day-ahead prices
over a full held-out calendar year, applying a common pre-auction information
boundary and rolling-origin design to linear quantile regression, a quantile
regression forest (QRF), a LEAR-QRA benchmark and an encoder-decoder quantile
long short-term memory (LSTM) network. Two routes to more reliable intervals are
then compared: conformal recalibration and Quantile Regression Averaging (QRA).
Aggregate scores separate the LSTM and QRF only weakly, with a non-significant
4.4\% difference in continuous ranked probability score (CRPS), but their
conditional behaviour differs sharply. Above 200 GBP/MWh the LSTM covered 0.12
of outcomes and QRF 0.71, consistent with the forest retaining realised spikes
in its leaf distributions while the network learned a smooth upper tail.
Conformal recalibration corrected marginal coverage but did not improve the
interval score or materially repair spike coverage. QRA combining the LSTM, QRF
and LEAR-QRA forecasts reduced CRPS to 5.14 GBP/MWh and raised coverage above
200 GBP/MWh to 0.65 at unchanged interval width. Recalibration and combination
therefore address different failures and are not substitutes.
\end{abstract}

\begin{keyword}
Probabilistic forecasting \sep Electricity price forecasting \sep
Quantile regression averaging \sep Conformal prediction \sep Day-ahead market
\sep Tail reliability
\end{keyword}

\end{frontmatter}

%==============================================================================
\section*{Nomenclature}
\noindent
\begin{tabular}{@{}ll@{\hspace{2.4em}}ll@{}}
ARX    & Autoregressive with exogenous inputs      & MPIL  & Mean prediction interval length \\
CQR    & Conformalised quantile regression         & OPMR  & Operational Planning Margin Report \\
CRPS   & Continuous ranked probability score       & PIT   & Probability integral transform \\
DM     & Diebold--Mariano                          & QR    & Quantile regression \\
EPEX   & European Power Exchange                   & QRA   & Quantile regression averaging \\
EPF    & Electricity price forecasting             & QRF   & Quantile regression forest \\
GB     & Great Britain                             & REMIT & Regulation on Energy Market \\
LEAR   & Lasso estimated autoregressive            &       & \quad Integrity and Transparency \\
LSTM   & Long short-term memory                    & RNN   & Recurrent neural network \\
MAE    & Mean absolute error                       & UK    & United Kingdom \\
\end{tabular}

%==============================================================================
\section{Introduction}
\label{sec:introduction}

Participants in the Great Britain (GB) wholesale electricity market use
day-ahead price forecasts to hedge volumetric exposure, schedule thermal and
flexible assets, and bid storage and demand-side response into spot auctions.
In the EPEX SPOT GB day-ahead auction, buy and sell orders for each delivery
hour of the following day are aggregated into demand and supply curves whose
intersection sets the hourly clearing price \citep{epexspot2026basics}, an
auction that cleared 44.23 TWh in 2025 \citep{epexspot2026annualresults}.
Positions are committed before the order-book cut-off, when realised demand,
renewable generation and plant availability remain unknown. Price formation
makes that commitment difficult. The clearing price reflects the cost of the
marginal plant, gas-fired generation set the GB wholesale price in 98\% of
hours in 2021 while producing just over 40\% of electricity
\citep{zakeri2023gas}, and renewables supplied 50.4\% of UK generation in 2024
\citep{desnz2025energytrends}. As the margin between available supply and
demand narrows, small changes in either can move the clearing price
disproportionately \citep{haben2021gb,sensfuss2008meritorder}.

\begin{figure}[htbp]
  \centering
  \includegraphics[width=\linewidth]{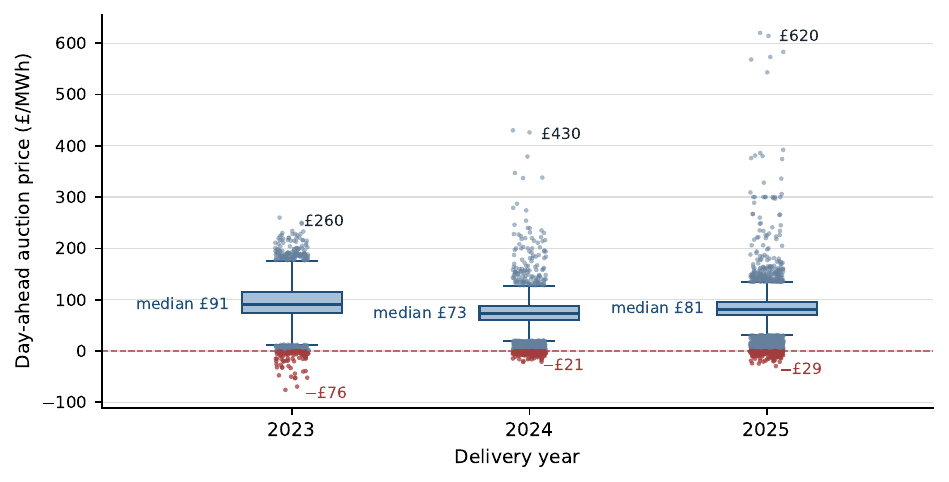}
  \caption[Distribution of hourly EPEX GB day-ahead auction prices by delivery
  year, 2023--2025.]{Distribution of hourly EPEX GB day-ahead auction prices by
  delivery year, 2023--2025, using EPEX SPOT data
  \citep{epexspot2026marketresults}. Median prices remain within a narrow band
  while the upper tail extends substantially, from an annual maximum of 260
  GBP/MWh in 2023 to 620 GBP/MWh in 2025.}
  \label{fig:intro_price}
\end{figure}

Figure~\ref{fig:intro_price} shows where that difficulty now concentrates.
Between 2023 and 2025 the annual median price stayed within a band of roughly
18 GBP/MWh, while the annual maximum rose from 260 to 620 GBP/MWh, hours above
200 GBP/MWh rose from 30 in 2024 to 52 in 2025, and negative-price hours rose
from 86 to 149 \citep{epexspot2026marketresults}. The centre of the
distribution is comparatively stable; the tails are not. For a retailer
carrying unhedged peak exposure, a battery optimiser deciding whether to commit
capacity day-ahead, or a flexible thermal generator, the central forecast sets
a baseline while the tails govern downside risk
\citep{krishnamurthy2018storage}. Probabilistic forecasts are the natural
response, and the established principle is to maximise sharpness subject to
calibration \citep{gneiting2007jrssb}. The scores used to judge probabilistic forecasts, however,
are averages over all delivery hours. A model can attain near-nominal marginal
coverage and a competitive continuous ranked probability score (CRPS) while
failing systematically in precisely the regimes that
Figure~\ref{fig:intro_price} identifies as growing.

This paper develops a probabilistic forecasting methodology for hourly EPEX GB
day-ahead prices addressing that failure. Two model classes with opposite
tail behaviour are combined: a quantile regression forest, which derives each
quantile from realised training prices and therefore retains observed spikes
\citep{meinshausen2006qrf}, and an encoder-decoder quantile long short-term
memory network, which learns a smooth conditional distribution and is more
accurate through the centre. Combining them by Quantile Regression Averaging
\citep{nowotarski2015qra} over a full held-out calendar year, under a common
pre-auction information boundary and rolling-origin design, reduced CRPS by
19.8\% and the Winkler interval score by 38.1\% relative to a standard LEAR
benchmark \citep{lago2021benchmark} evaluated on identical data, while raising
coverage of the 90\% interval above 200 GBP/MWh from 0.06 to 0.65. Adding the
LEAR forecasts as a third constituent improved central accuracy further, taking
the CRPS reduction to 21.2\%, but left the tail results unchanged. Robustness
in the tails is therefore produced by the contrast between the two model
classes rather than by the size of the model pool.

The alternative approach is to repair the interval directly. Conformal
prediction uses forecast errors from held-out observations to move interval
bounds towards a target coverage level without retraining the underlying model
\citep{vovk2005algorithmic,romano2019cqr}. In this application we find while restoring marginal coverage to nominal it did not repair the tail.
Because the corrections are quantiles of pooled conformity errors, a single
adjustment is applied to every delivery hour, relocating the bounds without
indicating when a spike is likely, and coverage above 200 GBP/MWh rose only to
0.19. Recalibration and combination therefore address different failures and
are not substitutes, a distinction that has not, to the authors' knowledge,
been tested directly under a single evaluation framework.

Two further results follow from the same evaluation. The divergence between
marginal and conditional reliability is not confined to unconventional
architectures. The LEAR benchmark recorded the closest marginal coverage to
nominal of any model assessed, at 0.893, and covered 0.06 of hours above 200
GBP/MWh, so a forecast can satisfy the calibration diagnostic most commonly
reported and still fail almost completely where it matters most. Grouped
information ablation supports the structural account of why the two model
classes diverge, showing that they draw on different predictor families.

These results are established in the first probabilistic evaluation of GB
day-ahead prices conducted under an explicit pre-auction information boundary
over a full held-out calendar year, and the first, among the studies reviewed
here, to include a quantile regression forest or conformal prediction in this
market. The paper therefore contributes a structural account of why two leading
model classes fail in opposite tails, a controlled comparison of recalibration
against combination as remedies for an unreliable interval, and evidence that
aggregate rankings are uninformative about price-spike reliability.

Section~\ref{sec:literature} reviews relevant literature in probabilistic forecasting of electricity prices with a particular focus on the two models used here and evaluation frameworks.
Section~\ref{sec:data} defines the forecasting task and data,
Section~\ref{sec:methods} sets out the models and evaluation framework,
Section~\ref{sec:results} reports the held-out results, and
Sections~\ref{sec:discussion} and~\ref{sec:conclusion} discuss and conclude.

%==============================================================================
\section{Related work}
The field of probabilistic electricity price forecasting is vast; here we restrict our review to the models and evaluation frameworks used in this paper but for a complete review of the field we refer the reader to \citep{nowotarski2018review} and \citep{oconnor2025review}.
\label{sec:literature}
\subsection{Quantile benchmarks and forecast combination}

Probabilistic electricity price forecasting spans full-density models, scenario
methods and direct quantile approaches, but direct quantile modelling has become
one of the main practical routes for day-ahead prices
\citep{nowotarski2018review,oconnor2025review}. Quantile regression estimates a
specified conditional quantile of the price distribution directly, rather than
the conditional mean targeted by ordinary least squares, so a set of such
regressions traces out the whole conditional distribution
\citep{koenker1978regression}. Quantile methods are attractive because they
avoid imposing a fixed price distribution while still producing intervals that
can be assessed for calibration and sharpness.

Within this line, the Global Energy Forecasting Competition 2014 ranked methods
on a shared dataset under a common quantile-loss metric, and the top two teams
in its price track used QRA variants, establishing forecast combination as a
competitive standard for probabilistic EPF \citep{hong2016gefcom}. Later work
mainly improved how these combinations are selected and regularised.
\citet{uniejewski2021lqra} apply regularised QRA to large pools of point
forecasts, reporting pinball score gains of 2.1\% on the Polish market and
approximately 3\% on Nord Pool over the best non-regularised benchmark. Factor
Quantile Regression Averaging instead compresses the candidate forecasts through
latent factors; using British APX prices from July 2010 to December 2012,
\citet{maciejowska2016fqra} find that it delivers coverage closest to nominal at
both interval levels, reaching 46.7\% and 86.8\% against nominal 50\% and 90\%,
and records the best 90\% Winkler score at 25.11 against 25.40 for QRA and 27.66
for an ARX benchmark. Extending QRA to pool probabilistic quantiles directly,
\citet{cornell2024nem} report lower pinball loss at every assessed quantile and
lower CRPS than all constituents, while \citet{oconnor2025conformal} find QRA
efficient in Irish electricity markets. Under-coverage nevertheless persists
across these studies, showing that QRA performance depends on both constituent
quality and combination design. \citet{maciejowska2024hybridqra} find that a
standardised hybrid factor-QRA approach provides more reliable interval coverage
than conventional QRA on German and Polish prices, and
\citet{lipiecki2024postprocessing} find that combining QRA with conformal
prediction and isotonic distributional regression outperforms a distributional
neural network in German and Spanish markets.

Direct linear quantile regression sits beside QRA as a separate benchmark route,
offering a transparent view of how the conditional price distribution changes
with the predictors. Using half-hourly GB APX data from 2005 to 2012,
\citet{hagfors2016uk} estimate how fuel prices, lagged prices, demand and
reserve-margin forecasts relate to the distribution across nine quantiles.
\citet{osone2025ukqr} apply the same model class to a weighted average of Nord
Pool and EPEX SPOT GB day-ahead auction prices and report a mean pinball loss of
4.36 GBP/MWh for the median forecast over a 227-day test in 2023; the target
pools two GB auctions, so the result is contextual rather than directly
comparable to EPEX-only forecasts. Regularisation provides a further reference
when the input set contains many correlated predictors
\citep{uniejewski2016varsel}. The LEAR model and open benchmark framework of
\citet{lago2021benchmark} provide the standard high-dimensional autoregressive
reference implementation. Since LEAR produces point forecasts, probabilistic
LEAR benchmarks can apply QRA to models estimated over different calibration
windows, allowing their forecasts to contribute differently across the
conditional distribution.

\subsection{Quantile regression forests}

The quantile regression forest, introduced by \citet{meinshausen2006qrf},
extends random forests from conditional means to conditional distributions by
deriving each quantile from the forest weights assigned to training-set
observations. \citet{cornell2024nem} describe their South Australian application
as the first use of QRF for electricity price forecasting. The construction
allows a single fitted ensemble to capture nonlinear predictor interactions and
changing conditional spread without imposing a parametric price distribution,
and producing all forecast quantiles in one fit keeps repeated rolling-origin
estimation practical. The same weighting structure, however, limits what the
forest can forecast: each QRF quantile is obtained from a weighted empirical
distribution of training responses and therefore cannot extrapolate beyond their
support \citep{meinshausen2006qrf}. This property is central to the present
study, because it implies that the forest's upper quantiles are bounded by
spikes that have already occurred, which is a liability for unprecedented events
and an advantage for recurrence of observed ones.

Outside a small standalone literature, forests appear mainly within conformal
EPF, either as base forecasters \citep{oconnor2025conformal} or as components of
conformal ensembles \citep{jensen2024encqr}. In the closest direct comparison,
\citet{cornell2024nem} find that QRF achieves the lowest pinball loss at
quantile levels from 0.90 to 0.975 on South Australian five-minute prices, while
every model remains under-covered against the nominal target. QRF performance
therefore cannot be judged by pinball loss alone, and its value for recent GB
day-ahead prices remains uncertain once coverage and tail reliability are
treated as central criteria.

\subsection{Recurrent networks for electricity price forecasting}

Recurrent neural networks pass information through an ordered input sequence
using a recurrent hidden state and can therefore represent temporal dependence
that may not be fully captured by a fixed set of engineered lags
\citep{elman1990structure}, while long short-term memory networks use gated
memory cells to retain or discard information over longer sequences
\citep{hochreiter1997lstm}. Applied to electricity prices, the evidence for
recurrence is mixed across markets and benchmark designs: gated recurrent units
outperform feed-forward networks on Turkish day-ahead prices
\citep{ugurlu2018rnn}, an LSTM is beaten by a two-stage vector autoregression on
German prices \citep{lehna2022comparison}, and a stand-alone recurrent network
beats a feed-forward deep network on more recent German-Luxembourg prices
\citep{benamor2025rnnlinear}.

Recent GB evidence is more directly relevant. \citet{cooke2025ukepf} evaluates
LSTM and GRU forecasts using half-hourly GB day-ahead data from January 2023 to
August 2025 and five rolling-origin validation folds, reporting an MAE of 14.26
GBP/MWh for the LSTM against 9.07 GBP/MWh for LightGBM. Recurrent sequence
modelling there does not improve upon tree-based models supplied with engineered
temporal and market features, although that evaluation is point-only and does
not assess the wider predictive distribution. Recurrent models are extended to
probabilistic EPF either by parameterising a predictive distribution or by
estimating conditional quantiles directly using pinball loss
\citep{liu2023lstnetquantile}; \citet{trebbien2023lstm} take the former route on
German-Luxembourg prices. In the closest GB study,
\citet{agakishiev2025multivariate} evaluate an LSTM on hourly EPEX day-ahead
prices over 196 out-of-sample days in 2022 and report a CRPS of 16.78 GBP/MWh
against 10.95 GBP/MWh for LASSO quantile regression. Recurrent structure alone
therefore does not guarantee an improvement over alternative probabilistic
models.

\subsection{Conformal prediction}

Producing probabilistic forecasts through parameterised distributions or
pinball-loss-trained quantiles does not ensure that prediction intervals attain
their nominal coverage. Conformal prediction addresses this reliability problem
by using forecast errors from held-out observations to adjust subsequent
intervals towards a target marginal coverage level \citep{vovk2005algorithmic}.
Conformalised quantile regression (CQR) applies the procedure to estimated lower
and upper quantiles, so that a quantile of the conformity scores determines the
adjustment to both bounds while retaining the nonlinear and asymmetric structure
learned by the underlying model \citep{romano2019cqr}.

\citet{kath2021conformal} provide an early direct application to electricity
prices. Their normalised conformal method scales forecast errors using an
estimate of local uncertainty, and its nominal 90\% intervals achieve empirical
coverage between 89.5\% and 90.3\% while QRA attains 80.9\%. The conformal
intervals are wider than the QRA intervals, demonstrating that improved coverage
must be considered together with the resulting loss of sharpness. Applying
conformal prediction to recurrent quantile forecasts,
\citet{chen2025oanqr} report seasonal coverage between 90.48\% and 99.11\% at
the 90\% nominal level on Singapore day-ahead prices, avoiding under-coverage
but sometimes materially exceeding the target.

Static conformal prediction uses a fixed calibration sample and typically
assumes exchangeability, an assumption strained by dependent and non-stationary
electricity-price series \citep{barber2023nexcp}. The resulting guarantee is
marginal: it applies on average across future observations rather than
separately to every delivery hour, price regime or tail, so a nominally
calibrated interval can still miss systematically under particular market
conditions \citep{romano2019cqr,barber2023nexcp}. Later work has addressed
temporal change through rolling conformity scores and feedback-controlled
corrections \citep{gibbs2021aci,angelopoulos2023conformalpid}.
\citet{jensen2024encqr} combine rolling scores with recurrent forecasts, raising
raw coverage values of 0.524, 0.550 and 0.724 to the nominal 0.900 level, and
\citet{brusaferri2025conformal} recover 80\% coverage from about 74\% under
temporal shift in European day-ahead markets using proportional-integral
feedback control. On Irish day-ahead and balancing prices,
\citet{oconnor2025conformal} find that an ensemble averaging quantile-regression
and time-series conformal bounds achieves the lowest average day-ahead Winkler
score at 38.37, narrowly ahead of the strongest standalone conformal methods at
38.43 and 38.55.

These results establish that conformal calibration can recover a target coverage
level under temporal change. What they do not establish is whether recovering
marginal coverage improves the interval where it matters, and that question
motivates the direct comparison with combination undertaken here.

\subsection{Evaluating probabilistic forecasts}

\citet{gneiting2007jasa} provide the formal basis for comparison through
strictly proper scoring rules, while \citet{gneiting2007jrssb} establish the
applied principle of maximising sharpness subject to calibration. This
distinction matters for the methods reviewed above, since conformal approaches
may improve interval validity by widening intervals while neural or
forest-based models may improve sharpness without being reliable in the tails.
Aggregate calibration diagnostics also have limits: a flat probability integral
transform (PIT) histogram is necessary but not sufficient for reliability,
because it can hide compensating conditional errors
\citep{diebold1998pit,hamill2001rankhist}, so marginal reliability must be
supplemented by conditional coverage checks across delivery hours, price regimes
and tails \citep{allen2025conditionalinterval}. At the same time, evaluating
models only on realised extremes can reward uninformative forecasts, the
forecaster's dilemma identified by \citet{lerch2017dilemma}, so tail evaluation
must be explicit but must not condition only on events known after the fact.
Metric choice should also reflect the forecast type and user need
\citep{messner2020windeval}, which for day-ahead prices means measuring both
general probabilistic accuracy and the reliability of extreme quantiles. For
flexible assets the relevant decision depends on the joint daily price path
\citep{krishnamurthy2018storage}, making multivariate path calibration the
fuller decision-aligned evaluation problem, though a more demanding one
\citep{bjerregaard2021mvreview}. A recurring caution is that headline scores are
comparable only under a shared evaluation framework, since differences in data,
test windows and significance testing can reverse apparent rankings
\citep{diebold1995comparing,lago2021benchmark}.

The empirical settings, methods and predictive findings of the most closely
related studies are summarised in Table~\ref{tab:lit_comparators}. Across the GB
literature, probabilistic forecasts are produced through different routes, and
although individual studies report calibration results, differences in targets,
base models and evaluation windows prevent comparison on a common basis. Recent
GB recurrent evidence is either point-only or lacks conformal recalibration,
while neither QRF nor conformal EPF has a published GB day-ahead application
among the studies reviewed. These gaps motivate the common evaluation framework
adopted here.

\begin{table}[!htb]
\centering
\footnotesize
\setlength{\tabcolsep}{2.8pt}
\renewcommand{\arraystretch}{1.16}
\caption{Reported out-of-sample results from the principal empirical comparator
studies. DA denotes day-ahead.}
\label{tab:lit_comparators}
\begin{tabular}{@{}>{\raggedright\arraybackslash}p{0.15\linewidth} >{\raggedright\arraybackslash}p{0.19\linewidth} >{\raggedright\arraybackslash}p{0.22\linewidth} >{\raggedright\arraybackslash}p{0.22\linewidth} >{\raggedright\arraybackslash}p{0.16\linewidth}@{}}
\toprule
Study & Market and test period & Method & Reported result & Relevance and limitation \\
\midrule
\citet{haben2021gb} & GB DA, hourly; test 1 January--17 August 2019 & Bid-stack X-model, ARX and ARIMAX with bootstrapped quantiles & Best mean pinball 1.71; the observed spike exceeded all 95th-percentile forecasts. & GB probabilistic benchmark. \\
\citet{osone2025ukqr} & UK DA, half-hourly; test 15 March--27 October 2023 & Linear QR using market fundamentals & Median pinball 4.36 GBP/MWh. & Recent UK QR; weighted two-auction target. \\
\citet{cornell2024nem} & South Australian spot, five-minute; test 2018--2021 & QRF and linear QR & QRF had the lowest pinball loss at $q=0.90$--$0.975$; all models under-covered. & Direct QRF evidence; non-GB market. \\
\citet{cooke2025ukepf} & GB DA, half-hourly; 1 January 2023--8 August 2025 & LSTM, GRU, LightGBM and XGBoost; five rolling-origin folds & LSTM MAE 14.26 GBP/MWh. & Recent GB recurrent comparison; point-only. \\
\citet{agakishiev2025multivariate} & GB EPEX/Nord Pool DA, hourly; test 1 March--12 September 2022 & Joint DMLP and LSTM distributions; LASSO QR & EPEX CRPS: LSTM 16.78 GBP/MWh; LASSO QR 10.95 GBP/MWh. & GB probabilistic comparison. \\
\citet{kath2021conformal} & Nord Pool DA, hourly; test 4 July--31 December 2013 & Lasso, KNN and SVM with normalised conformal prediction & 90\% conformal coverage 89.5--90.3\%, against 80.9\% for QRA. & Direct conformal EPF comparator. \\
\citet{oconnor2025conformal} & Irish I-SEM DA, hourly; data 2019--2022 & QR, QRA and conformal methods using LEAR, KNN, RF and LightGBM & Q-Ens recorded the lowest DA Winkler score of 38.37. & Closest conformal and QRA comparator; Irish market. \\
\bottomrule
\end{tabular}
\end{table}
\FloatBarrier

%==============================================================================
\section{Forecasting task, information boundary and data}
\label{sec:data}

\subsection{Task and information availability}
\label{sec:task}

The task is to produce probabilistic forecasts of hourly prices in the GB
day-ahead electricity market. The target is the EPEX SPOT GB hourly auction
price in GBP/MWh \citep{epexspot2026marketresults}. For each delivery day $D$,
the complete delivery-day price profile is forecast at the 09:20 Europe/London
order-book cut-off on $D-1$. The auction normally comprises 24 hourly products,
with 23 and 25 on the spring and autumn clock changes; the modelling grid
retains one row per local clock hour, so spring days contain 23 observations and
the two repeated autumn products are averaged when both are present.

For every delivery hour, each model produces a separate price forecast at each
of eleven quantile levels,
$\mathcal{T}=\{0.05,0.10,0.20,0.30,0.40,0.50,0.60,0.70,0.80,0.90,0.95\}$. The
0.50 quantile is the median forecast, while the 0.05 and 0.95 quantiles form the
bounds of the 90\% prediction interval and the 0.10 and 0.90 quantiles those of
the 80\% interval.

\begin{figure}[htbp]
    \centering
    \begin{subfigure}{\linewidth}
      \centering
      \includegraphics[width=0.84\linewidth]{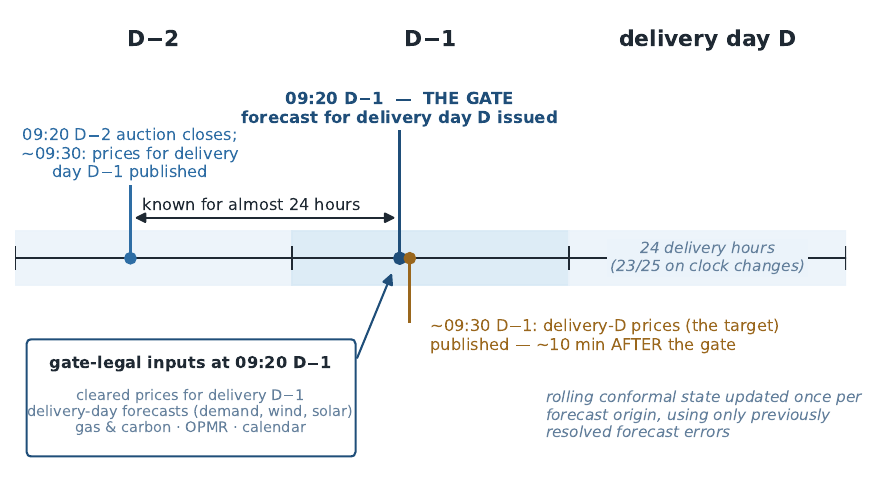}
      \caption{Forecast origin and information availability for delivery day $D$.}
    \end{subfigure}\\[6pt]
    \begin{subfigure}{\linewidth}
      \centering
      \includegraphics[width=\linewidth]{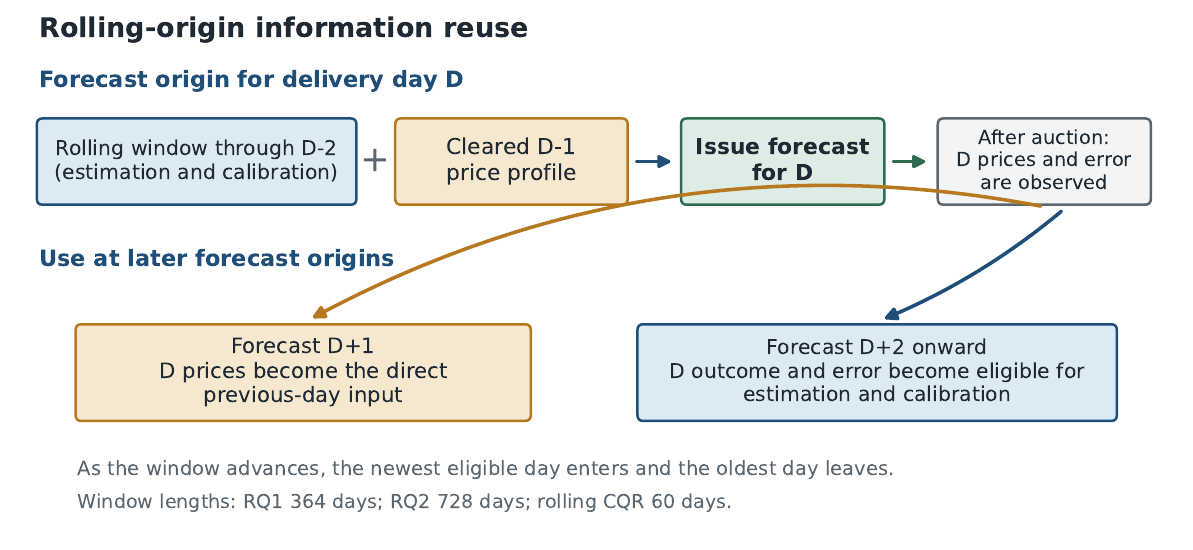}
      \caption{Information reuse under rolling-origin evaluation.}
    \end{subfigure}
    \caption[Forecast-time information boundary and rolling-origin information
    flow.]{Forecast-time information boundary and rolling-origin information
    flow. Only observations and vintages available to the left of the 09:20 gate
    enter the forecast. Newly cleared prices enter later forecasts only after
    publication, and realised outcomes and forecast errors enter rolling
    estimation and calibration with the stated $D-2$ endpoint.}
    \label{fig:method_gate}
\end{figure}

Figure~\ref{fig:method_gate}(a) formalises the information boundary applied to
every model. A predictor was admissible only if its realised value or valid
forecast vintage was available by the forecast origin. In particular, the
auction prices for delivery day $D-1$ had cleared on $D-2$, whereas the prices
for $D$ were not available until after the gate; training and conformal windows
therefore ended on $D-2$. Market operation and product timings follow the EPEX
auction documentation \citep{epexspot2026basics,epexspot2026tradingproducts}.
Figure~\ref{fig:method_gate}(b) shows how realised outcomes re-enter estimation
and calibration under the rolling-origin design described in
Section~\ref{sec:rolling}. Because published probabilistic EPF studies differ
considerably in how strictly forecast-time availability is enforced, and because
apparent skill is sensitive to that choice \citep{lago2021benchmark}, the
boundary is stated explicitly here and applied identically to every model,
including the benchmarks.

\FloatBarrier

\subsection{Data sources and sample}

The dataset spans 1 January 2023 to 31 December 2025 and contains 26,301 hourly
prices. Table~\ref{tab:target_summary} reports descriptive statistics by
delivery period. Prices were modelled in levels, since logarithms are invalid
for negative prices and an inverse hyperbolic-sine transformation did not
improve the validation forecasts. Negative prices and spikes were retained as
genuine market outcomes rather than trimmed or winsorised. The table shows that
price tails changed more markedly across the sample than the typical price
level: the negative-price share rose from 0.98\% to 1.70\% and the annual
maximum from 260 to 620 GBP/MWh, while the median moved within a band of
approximately 18 GBP/MWh.

\begin{table}[htbp]
\centering
\footnotesize
\setlength{\tabcolsep}{3pt}
\renewcommand{\arraystretch}{1.08}
\caption[Hourly EPEX GB day-ahead price summary by delivery period
(GBP/MWh).]{Hourly EPEX GB day-ahead price summary by delivery period
(GBP/MWh). Percentages are within-period shares. The 2025 calendar year is held
out for evaluation.}
\label{tab:target_summary}
\begin{tabular*}{\textwidth}{@{\extracolsep{\fill}}lrrrrrrrrrr@{}}
\toprule
Period & Hours & Mean & SD & Q1 & Median & Q3 & Min & Max & Negative & $>150$ \\
\midrule
2023 & 8,759 & 94.03 & 36.16 & 74.0 & 90.7 & 115.0 & $-75.7$ & 260 & 86 (0.98\%) & 537 (6.13\%) \\
2024 & 8,783 & 72.27 & 28.70 & 60.2 & 73.2 & 87.0 & $-21.3$ & 430 & 139 (1.58\%) & 78 (0.89\%) \\
2025 & 8,759 & 80.61 & 35.17 & 70.0 & 81.1 & 95.8 & $-29.4$ & 620 & 149 (1.70\%) & 123 (1.40\%) \\
Full sample & 26,301 & 82.29 & 34.68 & 67.0 & 81.0 & 98.0 & $-75.7$ & 620 & 374 (1.42\%) & 738 (2.81\%) \\
\bottomrule
\end{tabular*}
\end{table}

NESO supplied embedded-renewable forecasts, margin reports and interconnector
data \citep{neso2026opmr,neso2026dataportal}, while Elexon BMRS supplied
national-demand and transmission-wind forecast vintages alongside balancing,
generation and REMIT records \citep{elexon2026insights}. Daily gas and carbon
proxies completed the inputs \citep{investing2026ukgas,investing2026ukcarbon}.
For forecasts with multiple publications, the latest vintage available by the
gate was selected. Native hourly wind forecasts were preserved, half-hourly
continuous outturns were averaged to hours, and binary or event flags used the
hourly maximum. Auction-price predictors used the cleared $D-1$, $D-2$ and $D-7$
prices. Missing records were handled explicitly rather than by dropping
incomplete rows: after forecast-time alignment, the largest selected-feature
gaps are 177 balancing-lag hours (0.58\%), 175 seven-day interconnector hours,
172 seven-day price-lag hours and 96 commodity hours, most of them at the
beginning of the sample where the required earlier observations do not exist.
The tabular models replaced missing predictors with medians calculated
separately inside each active training window, and the linear model was
standardised using that window alone. For the fixed 24-position LSTM tensors,
isolated input gaps were interpolated within a delivery day and any still
incomplete training day was excluded; a spring clock-change day was padded only
for the model input, with forecasts scored on its 23 observed products. No price
target was imputed.

\subsection{Predictors}

Predictors were chosen to represent distinct drivers of day-ahead price
formation. Calendar terms capture recurring daily, weekly and annual patterns;
demand and the separate transmission-wind, embedded-wind and solar forecasts
describe the expected demand-renewables balance. Gas and carbon proxy the cost
of marginal thermal generation, while the Operational Planning Margin Report
(OPMR) and published REMIT events describe capacity margins and outage news.
Cleared auction prices capture price persistence, and lagged balancing,
generation and interconnector observations summarise the recent physical state
of the system. These choices are consistent with established GB and UK
price-forecasting specifications
\citep{hagfors2016uk,haben2021gb,osone2025ukqr}. Redundant alternatives were
removed through development-period comparisons rather than a mechanical
correlation threshold. Related predictors were retained where they represented
distinct physical channels, such as transmission and embedded wind, or
economically meaningful level, ratio and threshold forms, such as net demand,
renewable share and low residual load. In linear quantile regression, the
$\ell_1$ penalty limits coefficient redundancy.

Table~\ref{tab:feature_membership} groups the final predictor set, its source
and its forecast-time availability. All models shared the same pre-auction
information boundary but used model-appropriate representations. In particular,
the LSTM receives separate forecast channels rather than engineered balance
variables, allowing it to learn demand-renewable relationships directly.
Feature-level construction and model membership appear in
Table~\ref{tab:feature_membership_full}.

\begin{table}[htbp]
\centering
\footnotesize
\setlength{\tabcolsep}{4pt}
\renewcommand{\arraystretch}{1.14}
\caption[Grouped predictor set, source and forecast-time availability.]{Grouped
predictor set, source and forecast-time availability. Counts in parentheses
denote the number of tabular predictors and sum to 34.}
\label{tab:feature_membership}
\begin{tabular}{p{0.215\linewidth}p{0.285\linewidth}p{0.145\linewidth}p{0.255\linewidth}}
\toprule
Predictor group & Included predictors & Source & Availability at the 09:20 $D-1$ gate \\
\midrule
Calendar cycles and indicators (6) & Hour and month sine/cosine terms; day of week; bank-holiday flag & Calendar & Known in advance \\
Delivery-hour index (1) & Delivery hour & Calendar & Known in advance \\
Demand and renewable forecasts (4) & Demand; transmission wind; embedded wind; embedded solar & Elexon BMRS; NESO & Latest pre-gate delivery-$D$ forecast \\
Derived balance measures (4) & Net demand; renewable share; low-residual-load level and flag ($<5$ GW) & Derived & From the same eligible forecasts \\
Commodities (2) & Natural-gas and carbon prices & Investing.com & Latest eligible close, $\leq D-2$ \\
Normalised margin (1) & Normalised OPMR surplus & NESO & Latest valid OPMR, $\leq D-2$ \\
Margin level and revision (2) & OPMR national surplus; seven-day revision & NESO & Latest valid OPMR, $\leq D-2$ \\
Outage news (2) & REMIT published-event counts over one and seven days & Elexon BMRS & Published by the gate \\
Cleared auction prices (3) & Same-hour prices for delivery $D-1$, $D-2$ and $D-7$ & EPEX SPOT & Cleared before the gate \\
Balancing state (4) & System sell price and NIV at $D-2$ and $D-7$ & Elexon BMRS & Published outturns, $D-2$ and $D-7$ \\
Generation mix (3) & CCGT, wind and nuclear at $D-2$ & Elexon BMRS & Published outturns, $D-2$ \\
Interconnectors (2) & Net flow at $D-2$ and $D-7$ & NESO & Published outturns, $D-2$ and $D-7$ \\
\bottomrule
\end{tabular}
\end{table}
\FloatBarrier

\subsection{Nonlinearity in price formation}

\begin{figure}[htbp]
    \centering
    \includegraphics[width=0.86\linewidth]{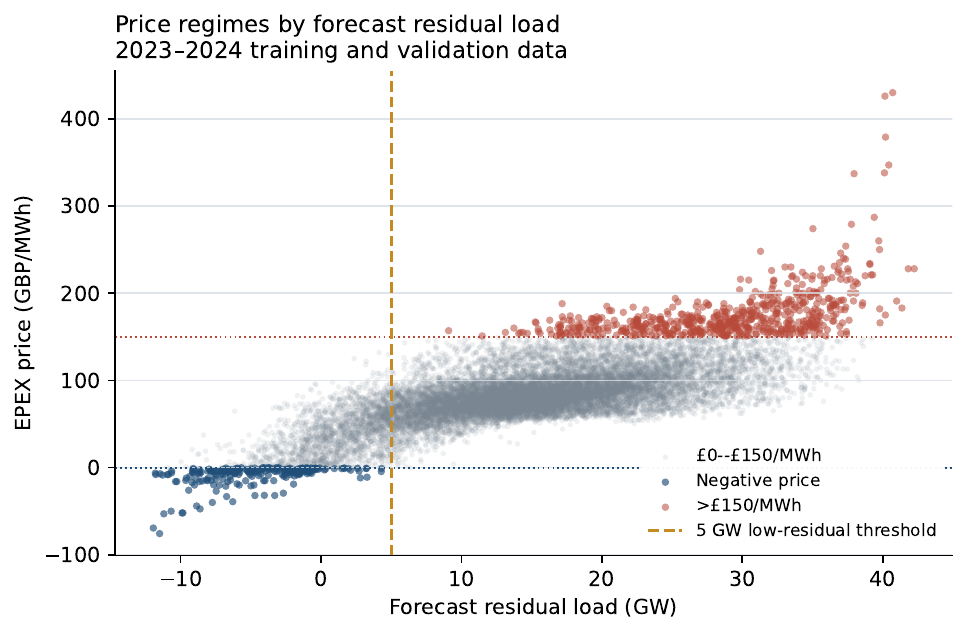}
    \caption[Prices against forecast residual load in the development
    data.]{Prices against forecast residual load in the 2023--2024 development
    data. Negative and high-price observations are highlighted, and the vertical
    line marks the 5 GW threshold used to define the low-residual-load flag.}
    \label{fig:method_residual}
\end{figure}

Figure~\ref{fig:method_residual} relates prices to untruncated forecast residual
load, defined as forecast demand less forecast wind and solar, using only the
2023--2024 development data. Negative prices occur only below 5 GW, where 10.1\%
of observations are negative, compared with none above that level. At residual
load of at least 30 GW, the median price is 136.5 GBP/MWh and 40.3\% of
observations exceed 150 GBP/MWh. The relationship between the supply-demand
balance and the price is therefore strongly nonlinear and asymmetric, with the
conditional spread widening sharply at high residual load. These patterns
motivate both the low-residual-load level and flag supplied to the tabular
models and the comparison between linear and nonlinear conditional quantile
estimators that follows.

\FloatBarrier

%==============================================================================
\section{Methods}
\label{sec:methods}

Model estimation used Python 3.13.5, with scikit-learn 1.7.2 and quantile-forest
1.4.1 for the tabular models and TensorFlow 2.21.0 with Keras 3.15.0 for the
recurrent models.

OpenAI Codex and Anthropic Claude assisted with code development and
verification. The lead author reviewed and validated all retained code and
results.

\subsection{Rolling-origin design}
\label{sec:rolling}

The data preceding the test period covered 1 January 2023 to 31 December 2024.
Initial fitting used 1 January 2023 to 30 June 2024 (13,126 hours over 547
delivery days), and validation used 1 July to 31 December 2024 (4,416 hours over
184 days). The held-out test covered 1 January to 31 December 2025 (8,759 hours
over 365 days). Validation informed feature, architecture and hyperparameter
decisions; the test did not.

Models were updated through rolling-origin evaluation rather than fitted once.
Once an earlier test outcome became available it could enter a later rolling
window, but never its own or an earlier forecast. This design follows
established electricity-price benchmark practice \citep{lago2021benchmark} and
discards older regimes as new observations arrive. The linear and forest models use 364
trailing delivery days, providing approximately one annual cycle while retaining
responsiveness to regime change, and the recurrent model uses 728 days because
the higher-variance estimator benefited from the larger sample in development
comparisons. The historical baseline was recomputed daily, and estimated tabular
and recurrent models were refitted every seven forecast days. Architecture and
hyperparameter comparisons were limited to targeted development alternatives; no
automated search was used. An expanding QRF, which retains every eligible
observation, was kept as a window-mode sensitivity.

\subsection{Evaluation framework}
\label{sec:evaluation}

A common evaluation framework is required because the models produce different
forms of probabilistic output and a ranking based on one statistic can conceal
poor calibration or excessively wide uncertainty. Models are therefore assessed
on the same observations and, where their outputs permit, the same quantile or
interval levels. The framework follows established probabilistic and
power-forecast evaluation practice
\citep{pinson2007windeval,nowotarski2018review}.

For an error $u=y-\widehat q_{\tau}$, the quantile or pinball loss is

\begin{equation}
\rho_{\tau}(u)=u\{\tau-\mathbb{I}(u<0)\}.
\label{eq:pinball}
\end{equation}

Its asymmetric weights vary with $\tau$, so its expected value is minimised by
the corresponding conditional quantile \citep{koenker1978regression}. Models
fitted directly at specified quantiles are trained using
Equation~\eqref{eq:pinball}, which is also used to evaluate all models at each
quantile. The headline score is CRPS, which integrates quantile loss across the
predictive distribution and jointly rewards calibration and sharpness
\citep{matheson1976crps}. It is approximated using the trapezoidal rule over the
eleven-quantile grid,

\begin{equation}
\widehat{\operatorname{CRPS}}_{11}=
2\sum_{j=1}^{10}\frac{
\rho_{\tau_j}(y-\widehat q_{\tau_j})+
\rho_{\tau_{j+1}}(y-\widehat q_{\tau_{j+1}})}{2}
(\tau_{j+1}-\tau_j),
\label{eq:crps_grid}
\end{equation}

and averaged across forecast hours, retaining GBP/MWh units. Since $\mathcal{T}$
spans 0.05 to 0.95, Equation~\eqref{eq:crps_grid} is a finite-grid approximation
rather than exact continuous CRPS.
Mean pinball loss is also reported separately at each quantile, since it
identifies whether differences in the CRPS ranking arise in the centre or the
tails. At $\tau=0.50$, pinball loss is one half of absolute error, and median
MAE is retained as a point-forecast reference.

Empirical coverage is the fraction of outcomes lying within an interval and mean
prediction interval length (MPIL) is its average width in GBP/MWh. For
observations $i=1,\ldots,N$,

\begin{equation}
\widehat{\operatorname{Cov}}(l,u)
=\frac{1}{N}\sum_{i=1}^{N}\mathbb{I}(l_i\leq y_i\leq u_i),
\qquad
\operatorname{MPIL}(l,u)=\frac{1}{N}\sum_{i=1}^{N}(u_i-l_i).
\end{equation}

Coverage and MPIL are reported together because interval width is meaningful
only relative to nominal coverage
\citep{christoffersen1998interval,gneiting2007jrssb,pinson2007windeval}. Both
are calculated for the 90\% interval $[\widehat q_{0.05},\widehat q_{0.95}]$ and
the 80\% interval $[\widehat q_{0.10},\widehat q_{0.90}]$, with 90\% designated
as the headline. The Winkler interval score combines width and coverage
\citep{winkler1972interval}:

\begin{equation}
S_{\alpha}(l,u;y)=(u-l)+\frac{2}{\alpha}(l-y)\mathbb{I}(y<l)
+\frac{2}{\alpha}(y-u)\mathbb{I}(y>u).
\end{equation}

When $y$ lies inside $[l,u]$ the score is the interval width; lower and upper
misses of equal distance receive the same penalty, and narrower intervals are
rewarded only when they maintain coverage.

Numeric scores are supplemented by two distributional diagnostics. A reliability
diagram plots each nominal quantile $\tau$ against the observed proportion
satisfying $y\leq\widehat q_{\tau}$, so that a calibrated forecast follows the
diagonal and the direction of deviation identifies over- or under-predicted
quantiles \citep{pinson2007windeval}. An approximate PIT locates each outcome
within the eleven-quantile distribution, and calibration implies an
approximately uniform histogram \citep{diebold1998pit}. The PIT is approximate
because it is interpolated between the eleven forecast quantiles, and apparent
uniformity is not treated as sufficient evidence of conditional calibration
\citep{hamill2001rankhist}.

Because aggregate diagnostics are the object of study here, conditional
performance is reported systematically rather than incidentally. Lower- and
upper-tail miss rates distinguish the direction of interval failure, and
conditional coverage is reported for negative prices and for high-price hours
above 150 GBP/MWh, with 200 GBP/MWh as a stricter tail check. Both thresholds
were fixed before the test period was examined, following the caution of
\citet{lerch2017dilemma} against conditioning only on events identified after
the fact.

Models are compared pairwise using the Diebold--Mariano test
\citep{diebold1995comparing} on daily total eleven-quantile CRPS, obtained by
summing the hourly losses across each delivery day. A one-sided test assesses
whether model A has significantly lower loss than model B, and Newey--West HAC
variance accounts for dependence between consecutive daily loss differences
\citep{neweywest1987hac}, using lag $\lfloor 1.5T^{1/3}\rfloor$ for $T$ days.

Finally, a grouped ablation assesses the contribution of information families
after the final specifications were fixed. Each ablation refits the same
leakage-safe rolling protocol over the test period after removing one predictor
group, and effects are measured by the deterioration in out-of-sample pinball
loss relative to the complete model.

\subsection{Benchmarks and tabular models}

The baseline is the empirical distribution of prices at the same delivery hour
over the trailing 364 days, recomputed at every forecast origin. Its eleven
empirical quantiles provide a persistence-based reference against which the
value of additional feature information is assessed. It is robust to negative
prices and makes no distributional assumption, but cannot condition on the
current demand, renewable or market state. Its role as an explicit naive
comparator follows electricity-price benchmark practice
\citep{lago2021benchmark}.

Linear quantile regression provides the estimated benchmark and was fitted
separately for each delivery hour and quantile using the 31 applicable
predictors in Table~\ref{tab:feature_membership}. Delivery hour and its sine and
cosine terms were omitted because they are constant within each hourly
regression. An $\ell_1$ penalty controlled the correlated predictor set,
following established practice \citep{uniejewski2016varsel,osone2025ukqr}.

The quantile regression forest provides the nonlinear tabular comparison and
used all 34 predictors. A single fitted forest returned all eleven quantiles.
Since these quantiles are drawn from weighted training responses, QRF cannot
forecast beyond the price support represented in its training leaves
\citep{meinshausen2006qrf}. The sequence of historical, linear and forest models
separates gains from additional predictors from gains due to nonlinear
flexibility. Where separately estimated quantiles crossed, the eleven outputs
were placed in ascending order within that forecast row before scoring,
following an established post-processing convention \citep{uniejewski2021lqra}.

LEAR supplies the high-dimensional autoregressive benchmark from the
\texttt{epftoolbox} implementation of \citet{lago2021benchmark}. Each of 24
hourly LASSO regressions used 96 price-lag predictors, demand and total
renewable forecasts at $D$, $D-1$ and $D-7$ for all 24 hours, and seven weekday
indicators. Continuous predictors and targets used the package's invariant
scaling, and the penalty was selected by LARS with the Akaike information
criterion before a coordinate-descent LASSO refit. Models were refitted weekly
under the common $D-2$ boundary at trailing windows of 364, 546 and 728 days.
The resulting three point forecasts were converted to eleven LEAR-QRA quantiles
by quantile regression at each level. A 273-day member was excluded from the
principal specification after its near-square estimation problem produced
material convergence instability; including it is reported as a sensitivity.
This canonical feature set is intentionally narrower than the study-specific
inputs used by the other models, preserving LEAR's role as a reference
benchmark rather than creating a feature-matched variant. On the shared
two-exogenous specification, the implementation reproduced the package scaler,
features, selected penalties, coefficients and forecasts exactly.

\subsection{Encoder–Decoder LSTM}

The recurrent model is a direct multi-horizon sequence-to-sequence quantile
forecaster, hereafter the Encoder–Decoder LSTM, following the direct
sequence-to-sequence quantile framework of \citet{wen2017mqrnn}. All horizons
are forecast directly, so predicted prices are not fed recursively into later
horizons. This avoids recursive error accumulation while allowing structure to
be shared across the day.

\begin{figure}[htbp]
    \centering
    \includegraphics[width=\linewidth]{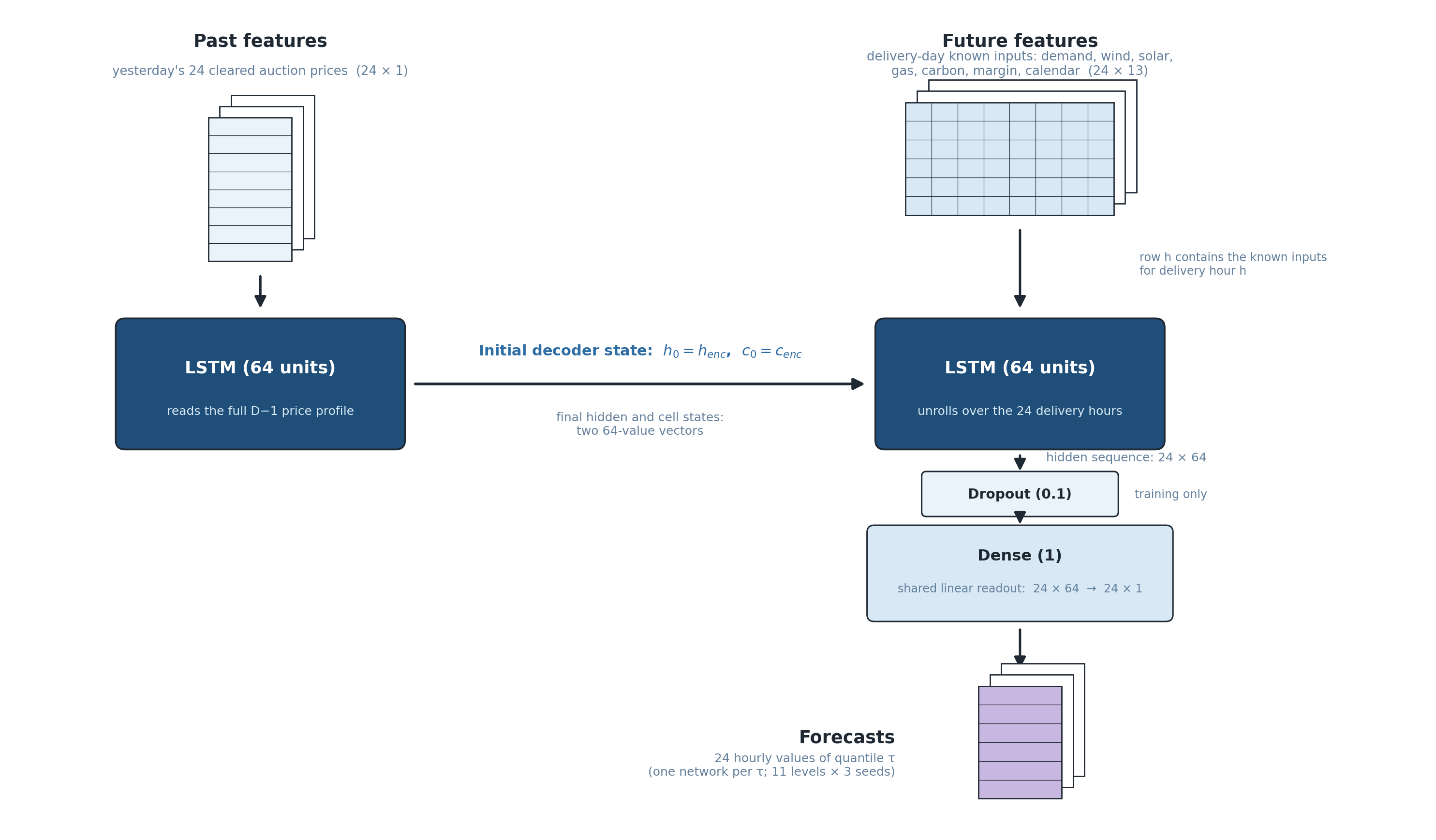}
    \caption[Encoder–Decoder LSTM architecture.]{Encoder–Decoder LSTM architecture. The
    encoder receives the $24\times1$ $D-1$ price profile. The decoder receives
    13 known-future channels for each of 24 hours and returns a $24\times1$
    forecast for one quantile. Separate networks cover eleven quantiles and
    three seeds.}
    \label{fig:method_lstm}
\end{figure}

As Figure~\ref{fig:method_lstm} shows, an encoder LSTM \citep{hochreiter1997lstm}
reads the full $24\times1$ vector of cleared prices for $D-1$. Its final hidden
and cell states initialise a decoder LSTM, which walks the 24 delivery hours
using a $24\times13$ matrix of known-future demand, transmission wind, embedded
wind, embedded solar, gas, carbon, normalised OPMR surplus, day of week, the
bank-holiday flag, and the sine and cosine transformations of delivery hour and
month. For each quantile level, the network produces the complete $24\times1$
delivery-day forecast in a single forward pass. Keeping realised history in the
encoder and known future conditions in the decoder makes the forecast-time
distinction explicit. Encoder-decoder networks have precedent in probabilistic
power-market forecasting \citep{bottieau2020imbalance}, and the 24-in/24-out
representation is consistent with horizon-sequence renewable forecasting
\citep{kean2026wind}.

Each network was trained using the pinball loss in
Equation~\eqref{eq:pinball}, which targets the corresponding conditional
quantile directly without imposing a parametric price distribution. Inputs and
targets were standardised using only the fitting portion of each rolling window,
and forecasts were transformed back to GBP/MWh. For each quantile, forecasts
from seeds 42, 7 and 101 were averaged. Training settings are reported in
Table~\ref{tab:lstm_training}. The quantile networks were estimated separately,
so monotone rearrangement was applied within each seed before the forecasts were
averaged, ensuring non-crossing final quantiles.

\subsection{Conformal recalibration}
\label{sec:cqr}

Conformal recalibration is applied to the LSTM, the strongest individual model
over the July to December 2024 validation period, to test whether coverage of
its 80\% and 90\% intervals could be improved during the held-out year without
sacrificing sharpness. Both variants use lower- and upper-bound errors from
earlier out-of-sample day-ahead forecasts, rather than training residuals, to
calculate how far those bounds should move. Static conformal prediction holds
the resulting corrections fixed, whereas rolling conformal prediction
recalculates them before each forecast day using the most recent eligible
errors. Any resulting difference can therefore be attributed to the conformal
step, which has precedent in day-ahead and intraday power markets
\citep{kath2021conformal}.

For a raw lower-upper pair $[\widehat q_{\alpha/2},\widehat q_{1-\alpha/2}]$,
the asymmetric lower and upper conformity errors are

\begin{equation}
e_i^{L}=\widehat q_{\alpha/2,i}-y_i,
\qquad e_i^{U}=y_i-\widehat q_{1-\alpha/2,i}.
\end{equation}

A positive lower error means that the observed price fell below the lower bound,
while a positive upper error means that it exceeded the upper bound. Keeping the
errors separate allows the two tails to receive different corrections. With $n$
eligible calibration observations, let $c_L$ and $c_U$ be their respective
higher empirical quantiles at level $\min\{1,(1-\alpha/2)(1+1/n)\}$. The
adjusted interval is

\begin{equation}
\widehat C_{1-\alpha}(x)=
[\widehat q_{\alpha/2}(x)-c_L,\
 \widehat q_{1-\alpha/2}(x)+c_U].
\end{equation}

Errors are pooled across delivery hours within each calibration window,
producing one lower and one upper correction for each interval at every update.
Corrections may be negative, so an over-wide raw interval can contract. After
both levels are adjusted, the 90\% bounds are widened only where necessary to
enclose the 80\% interval. This is an asymmetric implementation of
conformalised quantile regression
\citep{romano2019cqr,brusaferri2025conformal}.

The static corrections applied throughout 2025 are estimated from 1 July to 31
December 2024 (184 days; 4,416 hourly errors). At each daily forecast origin the
rolling corrections instead use the most recent $W$ delivery days ending $D-2$,
with candidate 30-, 60- and 90-day windows compared over the same development
period and the 60-day window selected by absolute coverage error and then
Winkler score. Since the static calibration block coincides with the
model-selection validation period, the static result is treated as a calibration
comparison rather than a fully independent validation, and the rolling variant
carries the substantive conclusion. Dependence and non-stationarity violate the
exchangeability behind standard conformal guarantees, and rolling calibration
addresses this only partially \citep{barber2023nexcp}; coverage is therefore
interpreted as the observed hourly average over the test period, not as a
guarantee for every delivery hour or market condition.

\subsection{Quantile Regression Averaging}

QRA combines frozen constituent forecasts; none of the underlying models is
retrained \citep{nowotarski2015qra}. At each quantile level, the constituent
forecasts at that level serve as predictors in a quantile regression estimated
at the same level. Thus the 95th-percentile constituent forecasts become the
predictors of a regression targeting the 95th percentile. The original
two-model specification combines the LSTM and QRF. The principal three-model
specification adds the corresponding LEAR-QRA quantile, so that every
constituent enters on the same probabilistic basis.

The two- and three-model combinations are re-estimated before each seven-day
block of 2025 using matched out-of-sample forecasts and observed prices from 1
July 2024 through $D-2$. The LEAR-QRA benchmark is constructed similarly, but
its three window-specific point forecasts are available from January 2024 and
therefore enter an expanding calibration pool from that date. Delivery hours
are pooled in every combination, and monotone rearrangement is applied to
prevent quantile crossing \citep{chernozhukov2010rearrangement}.

As a sensitivity, an $\ell_1$-regularised version of the three-model QRA was
estimated with $\alpha\in\{10^{-4},10^{-3},10^{-2},10^{-1}\}$. The penalty was
selected on November--December 2024 forecasts before test evaluation, giving
$\alpha=10^{-3}$. This comparison tests whether coefficient shrinkage improves
the wider combination without changing its constituent pool or forecast grid.

The QRA estimation sample begins in the period used to select the LSTM
architecture and the conformal window length. This does not contaminate the
held-out evaluation, because every forecast scored in 2025 is produced from
constituent models and combination weights estimated only on data ending $D-2$,
and because no 2025 outcome enters the weights used to forecast it. The overlap
nevertheless means that the combination weights are estimated partly on a period
in which the constituents were themselves selected, which should be borne in
mind when interpreting the combination's development-period advantage.

Table~\ref{tab:model_provenance} consolidates the final settings and their
development provenance.

\begin{table}[htbp]
\centering
\footnotesize
\setlength{\tabcolsep}{3pt}
\renewcommand{\arraystretch}{1.15}
\caption{Final model, window and update specifications, their development-only
provenance, and the focused alternatives assessed before test evaluation.}
\label{tab:model_provenance}
\begin{tabular}{p{0.13\linewidth}p{0.15\linewidth}p{0.08\linewidth}p{0.29\linewidth}p{0.24\linewidth}}
\toprule
Method & Data used & Update & Final specification & Provenance and alternatives \\
\midrule
Historical quantiles & 364-day rolling window ending $D-2$ & Daily & Same-hour empirical quantiles on $\mathcal{T}$ & Parsimonious probabilistic reference; no tuned parameters \\
Linear QR & 364-day rolling window ending $D-2$ & Weekly & Per-hour fits; $\ell_1$ penalty 0.01; training-window imputation and scaling & Per-hour against pooled form and penalty sensitivity assessed in development \\
LEAR-QRA & 364-, 546- and 728-day rolling LEAR windows ending $D-2$ & Weekly & Canonical 247-feature LEAR; AIC-selected LASSO; QRA across window forecasts & \texttt{epftoolbox} implementation; 273-day member retained as a numerical-stability sensitivity \\
QRF & 364-day rolling window ending $D-2$ & Weekly & 500 trees; leaf size 5; \texttt{max\_features}$=0.5$; random state 42 & Leaf size, feature fraction, window, cadence and rolling against expanding estimation assessed in development \\
Encoder–Decoder LSTM & 728-day rolling window ending $D-2$; final 14 days held out & Weekly & Training settings in Table~\ref{tab:lstm_training} & 24/48-hour price history, 32/64 units, 364/728 days, batch size and early stopping assessed in development \\
Static CQR & Forecast errors from July--December 2024 & Fixed through 2025 & Separate lower and upper corrections for the 80\% and 90\% intervals & Fixed calibration design based on CQR precedent \\
Rolling CQR & Forecast errors from 60 delivery days ending $D-2$ & Daily & Separate lower and upper corrections for the 80\% and 90\% intervals & $W\in\{30,60,90\}$ selected over July--December 2024 by coverage error, then Winkler score \\
QRA & Matched constituent forecasts from 1 July 2024 through $D-2$ & Weekly & Two-model LSTM+QRF and principal three-model LSTM+QRF+LEAR-QRA combinations & Regularised three-model QRA selected on November--December 2024 as a sensitivity \\
\bottomrule
\end{tabular}
\end{table}
\FloatBarrier

%==============================================================================
\section{Results}
\label{sec:results}

All results refer to the held-out 2025 calendar year, comprising 8,759 hourly
forecasts over 365 delivery days.

\subsection{Aggregate accuracy, sharpness and calibration}
\label{sec:res_aggregate}

Table~\ref{tab:headline} reports the headline results. All estimated models
improve substantially on the historical benchmark, whose CRPS of 15.55 GBP/MWh
reflects its inability to condition on the current market state. Because that
benchmark is so weak, percentage improvements measured against it are
uninformative for model selection, and linear quantile regression is treated as
the reference point in what follows.

\begin{table}[htbp]
\centering
\footnotesize
\setlength{\tabcolsep}{4pt}
\renewcommand{\arraystretch}{1.12}
\caption[Probabilistic performance over the 2025 held-out year.]{Probabilistic
performance over the 2025 held-out year (8,759 hours). Coverage, MPIL and
Winkler score refer to the 90\% interval. Conditional coverage columns report
the share of outcomes inside the 90\% interval within each price regime
($n=149$, $123$ and $52$ respectively). Lower CRPS, MAE, MPIL and Winkler scores
are better; coverage is better closer to the nominal 0.90.}
\label{tab:headline}
\begin{tabular*}{\textwidth}{@{\extracolsep{\fill}}lrrrrrrrr@{}}
\toprule
Model & CRPS & MAE & Cov. & MPIL & Winkler & Cov. & Cov. & Cov. \\
      & (GBP/MWh) & (GBP/MWh) & at 90\% & (GBP/MWh) & (GBP/MWh) & $<0$ & $>150$ & $>200$ \\
\midrule
Historical benchmark & 15.55 & 20.89 & 0.864 & 92.7 & 147.4 & 0.00 & 0.15 & 0.00 \\
Linear QR            & 6.89  & 9.68  & 0.823 & 36.8 & 67.6  & 0.72 & 0.44 & 0.08 \\
LEAR-QRA             & 6.52  & 8.72  & 0.893 & 37.4 & 71.2  & 0.48 & 0.28 & 0.06 \\
QRF                  & 5.85  & 8.38  & 0.917 & 36.6 & 47.8  & 0.70 & 0.76 & \textbf{0.71} \\
Encoder–Decoder LSTM       & 5.59  & 7.89  & 0.895 & 32.5 & 51.2  & \textbf{0.87} & 0.51 & 0.12 \\
Two-model QRA        & 5.23  & 7.45  & 0.893 & 32.8 & \textbf{44.1} & 0.81 & \textbf{0.79} & 0.65 \\
Three-model QRA      & \textbf{5.14} & \textbf{7.26} & \textbf{0.900} & 32.8 & \textbf{44.1} & 0.81 & \textbf{0.79} & 0.65 \\
\bottomrule
\end{tabular*}
\vspace{3pt}
\par\footnotesize One-sided Diebold--Mariano tests on daily total CRPS: QRF
against the historical benchmark $p<0.0001$ and against linear QR $p=0.0079$;
the Encoder–Decoder LSTM against QRF $p=0.1949$; two-model QRA against the Encoder–Decoder LSTM
$p=0.0408$ and against QRF $p<0.0001$; three-model against two-model QRA
$p=0.0050$, against the LSTM $p=0.0111$, and against QRF and LEAR-QRA
$p<0.0001$.
\end{table}

Among the estimated models, QRF recorded a CRPS of 5.85 GBP/MWh against 6.89 for
linear quantile regression, a reduction of 15.1\% that is statistically
significant ($p=0.0079$). The clearest difference lies in interval quality
rather than in width. The two models produced 90\% intervals of almost identical
average length, at 36.6 and 36.8 GBP/MWh, yet covered 91.7\% and 82.3\% of
outcomes respectively. Linear quantile regression is therefore under-dispersed
at the nominal level and QRF is over-dispersed; the Winkler score adjudicates
between the two failure directions, and at 47.8 against 67.6 it favours QRF by a
wide margin.

The Encoder–Decoder LSTM recorded the lowest CRPS of any individual model, at 5.59
GBP/MWh, 4.4\% below QRF, with MAE falling from 8.38 to 7.89 GBP/MWh. Its 90\%
interval was 11.2\% narrower, at 32.5 against 36.6 GBP/MWh, while coverage
remained closer to nominal at 0.895 against 0.917. QRF nevertheless recorded the
lower Winkler score, at 47.8 against 51.2, and the CRPS difference between the
two models was not statistically significant ($p=0.1949$). On aggregate
evidence alone, therefore, the LSTM and QRF are close to indistinguishable: the
two scores that are usually treated as decisive disagree about the ranking, and
the difference in the headline score does not survive a significance test.

LEAR-QRA recorded a CRPS of 6.52 GBP/MWh and MAE of 8.72 GBP/MWh. Its 90\%
interval was wider than those of the other estimated benchmarks, at 37.4
GBP/MWh, while coverage remained near nominal at 0.893. It therefore supplies a
recognisable autoregressive benchmark, although it was weaker than QRF and the
LSTM as a stand-alone probabilistic forecast on this test year.

Two-model QRA reduced CRPS to 5.23 GBP/MWh. Adding LEAR-QRA reduced it by a
further 1.8\% to 5.14 GBP/MWh and reduced MAE from 7.45 to 7.26 GBP/MWh. The
incremental CRPS reduction was statistically significant ($p=0.0050$), as were
the three-model forecast's advantages over the LSTM, QRF and LEAR-QRA. Coverage
was exactly nominal to the reported precision at 0.900, with MPIL of 32.8
GBP/MWh and a Winkler score of 44.1. The extra constituent therefore improved
the distributional score without widening the interval.

\begin{figure}[htbp]
  \centering
  \includegraphics[width=\textwidth]{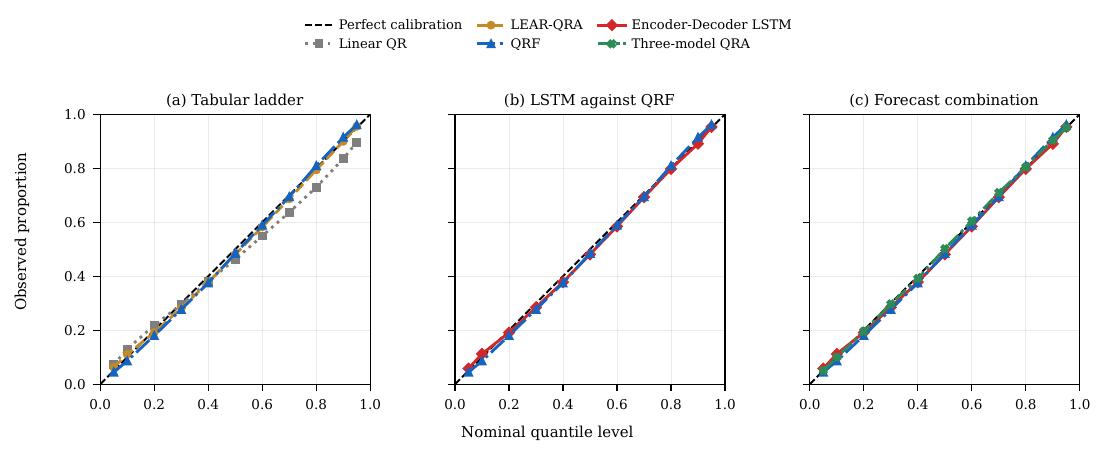}
  \caption[Quantile reliability across the model comparisons during 2025.]{Quantile
  reliability across the model comparisons during 2025: (a) the tabular ladder,
  (b) the Encoder–Decoder LSTM against QRF and (c) three-model QRA against its
  principal constituents. The diagonal denotes perfect empirical calibration.}
  \label{fig:reliability}
\end{figure}

Reliability diagrams for the tabular ladder, the LSTM against QRF and the
three-model QRA comparison are shown in Figure~\ref{fig:reliability}. The curves
lie close to the diagonal across the quantile grid, and the corresponding PIT
histograms are reported in Appendix~\ref{app:supplementary}. Marginal
calibration diagnostics of this kind do not separate the models, which is the
point developed in Section~\ref{sec:res_conditional}.

\FloatBarrier

\subsection{Conditional and tail performance}
\label{sec:res_conditional}

\begin{figure}[htbp]
  \centering
  \includegraphics[width=\textwidth]{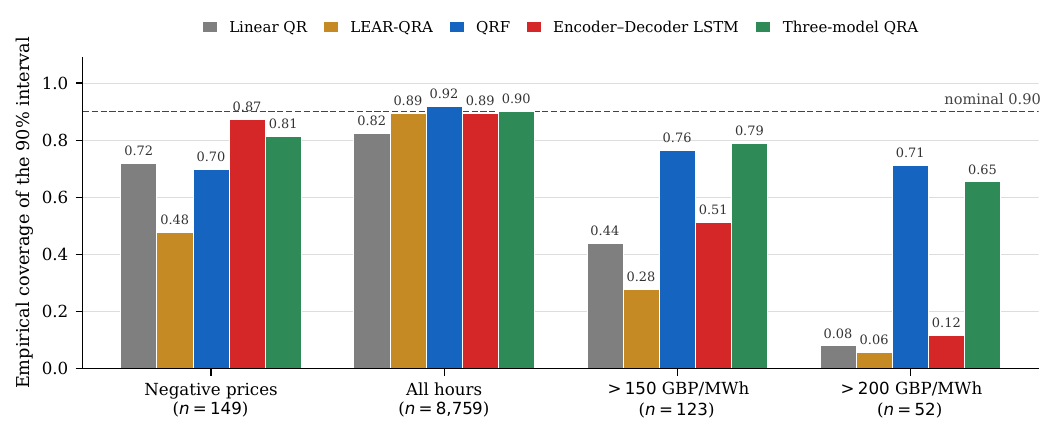}
  \caption[Conditional coverage of the 90\% interval by price regime during
  2025.]{Conditional coverage of the 90\% interval by price regime during 2025.
  Thresholds were fixed before the test period was examined. The ordering of the
  five estimated models above 200 GBP/MWh is close to the reverse of their
  ordering by aggregate CRPS.}
  \label{fig:cond_coverage}
\end{figure}

Figure~\ref{fig:cond_coverage} reports coverage of the 90\% interval within
price regimes fixed in advance. The regimes separate the models far more sharply
than the aggregate scores, and in the strictest regime they reorder them.

During negative-price hours the Encoder–Decoder LSTM covered 0.87 of outcomes, ahead
of both forecast combinations at 0.81, linear quantile regression at 0.72 and QRF
at 0.70. Above 150 GBP/MWh the ordering changes: both combinations covered
0.79 and QRF 0.76, while the LSTM
covered 0.51 and linear quantile regression 0.44. Above 200 GBP/MWh, across the
52 hours that exceeded that level, QRF covered 0.71 and both combinations
0.65, whereas the LSTM covered 0.12, LEAR-QRA 0.06 and linear quantile
regression 0.08.

The comparison between the LSTM and QRF is the substantive one. The model with
the lower CRPS, the lower MAE, the narrower interval and the closer-to-nominal
marginal coverage captured 12\% of the most extreme outcomes, against 71\% for
the model it appeared to beat. Neither the aggregate scores in
Table~\ref{tab:headline} nor the reliability diagram in
Figure~\ref{fig:reliability} anticipates this difference, and the
Diebold--Mariano test on daily CRPS treats the two models as statistically
indistinguishable.

Mean pinball loss by quantile in Table~\ref{tab:pinball} locates where the
differences arise. The LSTM recorded lower loss than QRF at every quantile from
0.10 to 0.80, with the largest absolute advantage at 0.40 (3.79 against 4.07
GBP/MWh), and the two recorded the same loss at 0.05 to the reported precision. QRF recorded lower loss at 0.90 (1.95
against 2.05) and at 0.95 (1.22 against 1.39). The two models are therefore not uniformly ranked: the
recurrent model is more accurate through the centre of the conditional
distribution and the forest is more accurate in the upper tail. Three-model QRA
recorded the lowest displayed loss from 0.20 to 0.90 and matched two-model QRA
to the reported precision in both tails. The divergence is starkest for the standard benchmark. LEAR-QRA attained marginal coverage of 0.893, closer to nominal than any other model assessed, yet covered 0.06 of the hours above 200 GBP/MWh. A forecast that is calibrated by the diagnostic most commonly reported was reliable in six per cent of the hours in which reliability matters most. 

\begin{table}[htbp]
\centering
\scriptsize
\setlength{\tabcolsep}{2.5pt}
\renewcommand{\arraystretch}{1.1}
\caption[Mean pinball loss at all eleven forecast quantiles during 2025.]{Mean
pinball loss in GBP/MWh at all eleven forecast quantiles during 2025. Lower
values are better. The best value in each column is shown in bold.}
\label{tab:pinball}
\begin{tabular*}{\textwidth}{@{\extracolsep{\fill}}lrrrrrrrrrrr@{}}
\toprule
Model & $q_{0.05}$ & $q_{0.10}$ & $q_{0.20}$ & $q_{0.30}$ & $q_{0.40}$ & $q_{0.50}$ & $q_{0.60}$ & $q_{0.70}$ & $q_{0.80}$ & $q_{0.90}$ & $q_{0.95}$ \\
\midrule
Historical benchmark & 3.52 & 5.97 & 8.72 & 9.81 & 10.28 & 10.44 & 10.24 & 9.53 & 8.14 & 5.65 & 3.85 \\
Linear QR & 1.32 & 2.11 & 3.27 & 4.10 & 4.61 & 4.84 & 4.76 & 4.46 & 3.84 & 2.85 & 2.06 \\
LEAR-QRA & 1.86 & 2.77 & 3.72 & 4.17 & 4.36 & 4.36 & 4.18 & 3.83 & 3.25 & 2.36 & 1.70 \\
QRF & 1.17 & 1.94 & 3.02 & 3.71 & 4.07 & 4.19 & 4.07 & 3.69 & 2.97 & 1.95 & 1.22 \\
Encoder–Decoder LSTM & 1.17 & 1.91 & 2.88 & 3.48 & 3.79 & 3.94 & 3.84 & 3.49 & 2.91 & 2.05 & 1.39 \\
Two-model QRA & \textbf{1.06} & \textbf{1.74} & 2.70 & 3.29 & 3.59 & 3.73 & 3.63 & 3.30 & 2.71 & 1.81 & \textbf{1.15} \\
Three-model QRA & \textbf{1.06} & \textbf{1.74} & \textbf{2.67} & \textbf{3.23} & \textbf{3.52} & \textbf{3.63} & \textbf{3.53} & \textbf{3.23} & \textbf{2.67} & \textbf{1.80} & \textbf{1.15} \\
\bottomrule
\end{tabular*}
\end{table}
\FloatBarrier

\subsection{Interval recalibration}
\label{sec:res_conformal}

Table~\ref{tab:conformal} reports the effect of conformal recalibration on the
LSTM intervals. At the 90\% level the raw intervals were already close to
nominal, at 0.895. Static conformal prediction raised coverage to 0.912 and
rolling conformal prediction to 0.900, while increasing MPIL from 32.5 to 34.3
and 34.1 GBP/MWh respectively. Neither improved the Winkler score, which moved
from 51.2 to 51.3 under static calibration and to 51.7 under rolling
calibration. The additional width therefore cost more than the additional
coverage returned.

\begin{table}[htbp]
\centering
\footnotesize
\setlength{\tabcolsep}{2.5pt}
\renewcommand{\arraystretch}{1.1}
\caption[Interval performance of the raw Encoder–Decoder LSTM and of static and
rolling conformal recalibration during 2025.]{Interval performance of the raw
Encoder–Decoder LSTM and of static and rolling conformal recalibration during
2025. Conditional columns report 90\% interval coverage for the 123 hours above
150 GBP/MWh and the 52 hours above 200 GBP/MWh.}
\label{tab:conformal}
\begin{tabular*}{\textwidth}{@{\extracolsep{\fill}}lrrrrrrrr@{}}
\toprule
Interval method & Cov. & MPIL 80\% & Winkler 80\% & Cov. & MPIL 90\% & Winkler 90\% & Cov. & Cov. \\
 & at 80\% & (GBP/MWh) & (GBP/MWh) & at 90\% & (GBP/MWh) & (GBP/MWh) & $>150$ & $>200$ \\
\midrule
Raw Encoder–Decoder LSTM & 0.778 & 23.0 & \textbf{39.5} & 0.895 & \textbf{32.5} & \textbf{51.2} & 0.512 & 0.115 \\
Static split CQR & 0.848 & 26.6 & 39.9 & 0.912 & 34.3 & 51.3 & 0.512 & 0.115 \\
Rolling asymmetric CQR & \textbf{0.801} & 24.6 & 39.8 & \textbf{0.900} & 34.1 & 51.7 & \textbf{0.553} & \textbf{0.192} \\
\bottomrule
\end{tabular*}
\end{table}

The distinction between the two variants was clearer at the 80\% level, where
the raw interval was genuinely under-covering at 0.778. Rolling calibration
raised coverage to 0.801, close to nominal, with less widening than static
calibration, which over-corrected to 0.848. Recent calibration errors were
therefore more representative than a fixed historical adjustment as forecast
errors changed through the year.

\begin{figure}[htbp]
  \centering
  \includegraphics[width=0.68\textwidth]{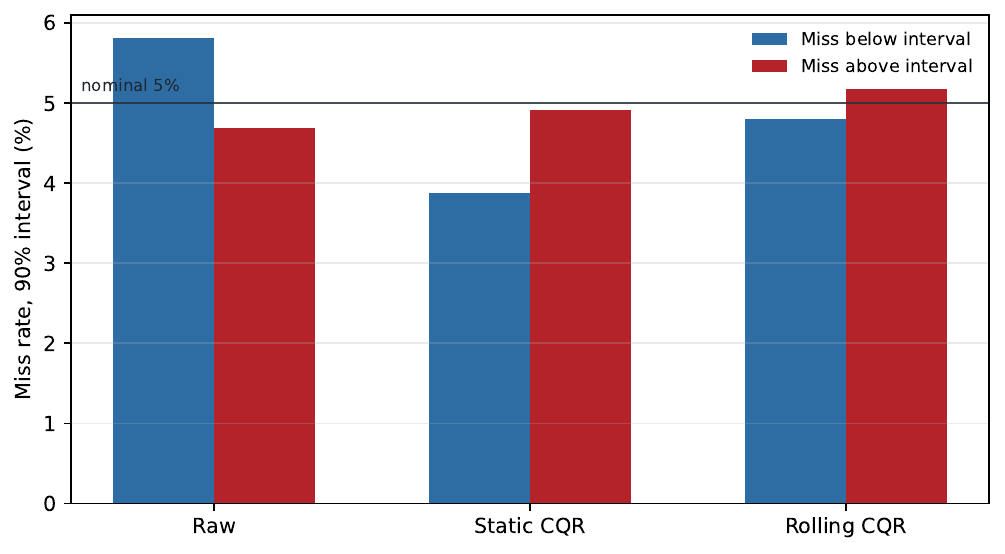}
  \caption[Lower- and upper-tail miss rates for the raw and conformal 90\%
  intervals during 2025.]{Lower- and upper-tail miss rates for the raw and
  conformal 90\% intervals during 2025. Each tail has a nominal miss rate of
  5\%.}
  \label{fig:cqr_tail_balance}
\end{figure}

Figure~\ref{fig:cqr_tail_balance} shows the direction of the remaining interval
failures. The raw interval missed 5.8\% of outcomes below the lower bound and
4.7\% above it, against a nominal 5\% in each tail. Static calibration moved
these to 3.9\% and 4.9\%, over-correcting the lower tail, while rolling
calibration produced 4.8\% and 5.2\%, the most balanced of the three.

Rolling conformal recalibration therefore did what it is designed to do: it
brought marginal coverage to the nominal level and balanced the two tails. What
it could not do is change which hours the interval covers. Because the
corrections are quantiles of pooled conformity errors, a single lower and a
single upper adjustment are applied to every delivery hour in the calibration
period, shifting all intervals by the same amount irrespective of the
conditions expected for that hour. The procedure relocates the bounds without
supplying information about when a spike is likely, so the upper-tail weakness
identified in Section~\ref{sec:res_conditional} is not the kind of failure that
recalibration addresses. The conditional results confirm this limitation.
Static recalibration left LSTM coverage unchanged at 0.512 above 150 GBP/MWh and
0.115 above 200 GBP/MWh. Rolling recalibration raised these only to 0.553 and
0.192, well below the corresponding QRF and QRA results.
\FloatBarrier

\subsection{Forecast combination}
\label{sec:res_qra}

Across the weekly two-model fits, LSTM coefficients ranged from 0.334 to 0.755
and QRF coefficients from 0.367 to 0.728. The pattern follows the conditional
strengths reported in Section~\ref{sec:res_conditional}: QRF received the larger
median coefficient in both tails, at quantile levels 0.05 to 0.20 and 0.90 to
0.95, while the LSTM received the larger coefficient across the centre, at
levels 0.30 to 0.80. The combination therefore reproduced, from matched
out-of-sample forecasts alone, the division of labour identified by the
conditional diagnostics.

The three-model combination preserved that structure while assigning LEAR-QRA
a smaller median coefficient of 0.176. Its median weight was approximately 0.10
at the 0.05, 0.90 and 0.95 quantiles and rose to 0.236--0.240 from the median to
the 0.70 quantile. This additional central information accounts for the CRPS
gain over two-model QRA without changing conditional spike coverage. The
regularised version was numerically indistinguishable, with CRPS 5.1382 against
5.1383 for the plain model. Although this minute difference was detectable in
the daily loss series ($p=0.037$), it changed no rounded metric, so the
unregularised specification is retained for parsimony.

\begin{figure}[htbp]
  \centering
  \captionsetup{skip=4pt}
  \includegraphics[width=\textwidth]{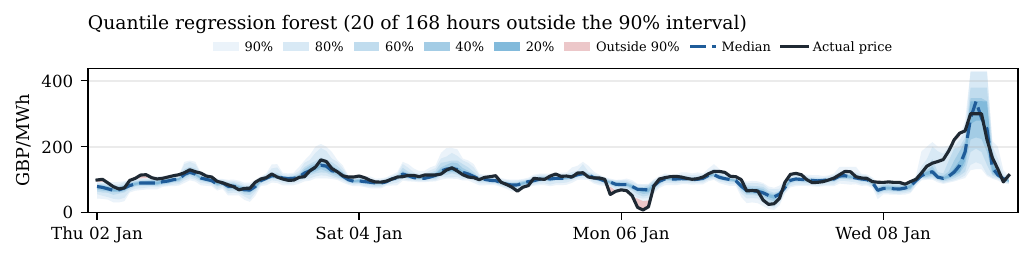}\par\vspace{-0.55em}
  \includegraphics[width=\textwidth]{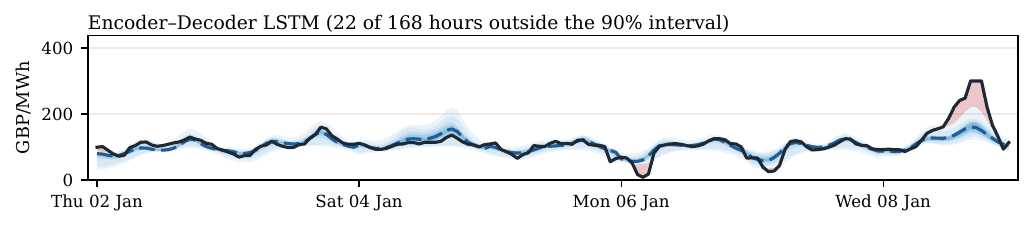}\par\vspace{-0.55em}
  \includegraphics[width=\textwidth]{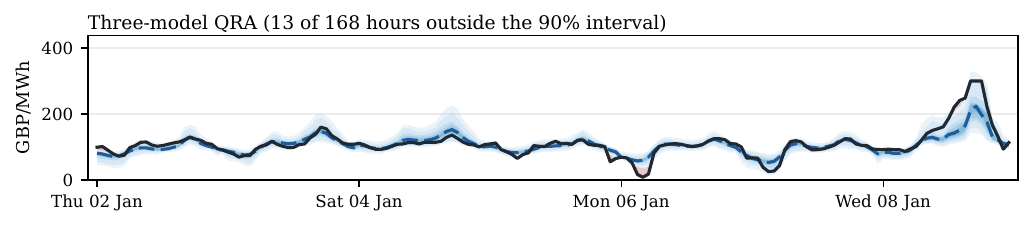}
  \caption[QRF, Encoder–Decoder LSTM and three-model QRA forecasts for 2--8 January 2025.]{QRF,
  Encoder–Decoder LSTM and three-model QRA forecasts for 2--8 January 2025. Red shading denotes
  excursions beyond the 90\% interval.}
  \label{fig:qra_case}
\end{figure}

Figure~\ref{fig:qra_case} illustrates the three-model combination over a week
containing a sustained high-price episode. Over 2 to 8 January 2025, it recorded a CRPS of
8.46 GBP/MWh against 10.73 for the LSTM and 9.15 for QRF. Three-model QRA and QRF each
covered 11 of the 13 hours above 150 GBP/MWh, compared with four for the LSTM,
while its 90\% interval averaged 51.2 GBP/MWh against 46.0 for the LSTM and
55.4 for QRF. The combined forecast tracked the LSTM through ordinary delivery
hours and moved towards QRF around the peaks.

\begin{figure}[htbp]
  \centering
  \includegraphics[width=\textwidth]{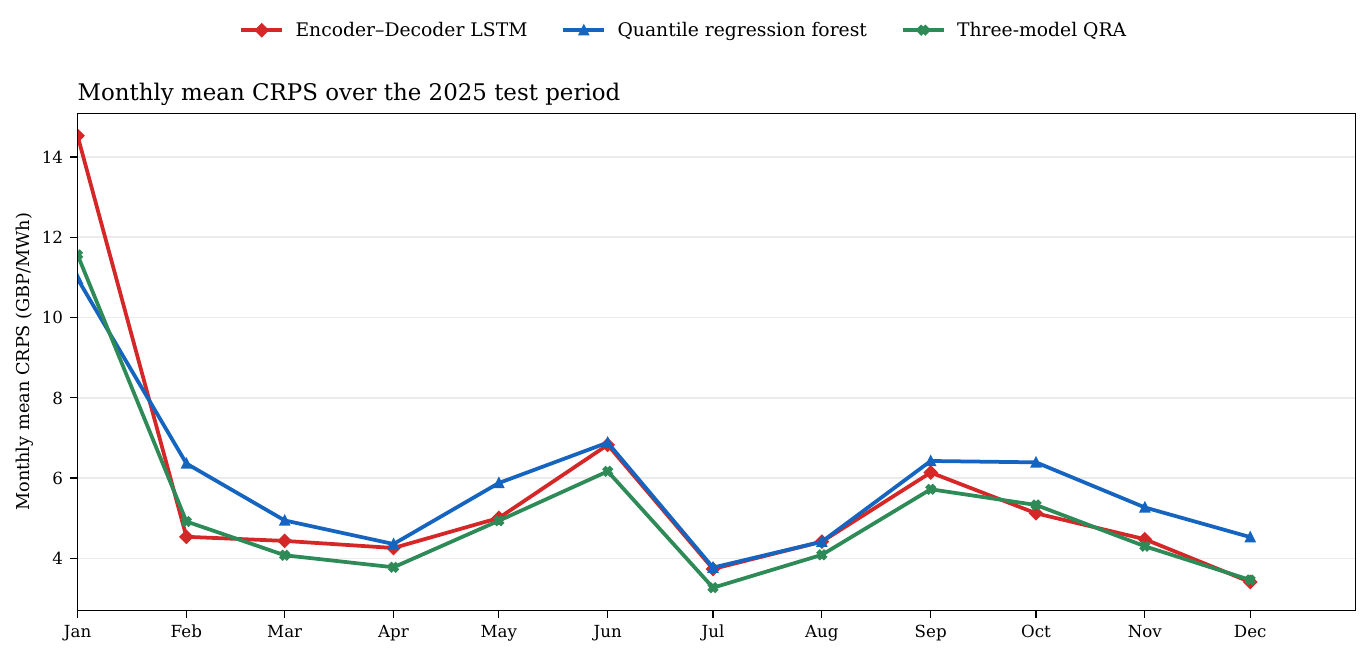}
  \caption{Monthly mean CRPS for three-model QRA and its principal constituent models during 2025.}
  \label{fig:monthly_crps}
\end{figure}

Monthly CRPS in Figure~\ref{fig:monthly_crps} shows that the three-model advantage
was not confined to a single episode, though neither was it uniform. Three-model QRA recorded the
lowest monthly CRPS in eight of the twelve months, with each principal constituent leading
in some of the remainder, and it recorded the lowest CRPS at 18 of the 24
delivery hours (Figure~\ref{fig:appendix_rq4_hourly_crps}). The largest monthly
losses for all three models occurred in January, when the high-price episodes
were concentrated.

\FloatBarrier

\subsection{Information content of the predictor families}
\label{sec:res_ablation}

\begin{figure}[htbp]
  \centering
  \includegraphics[width=\textwidth]{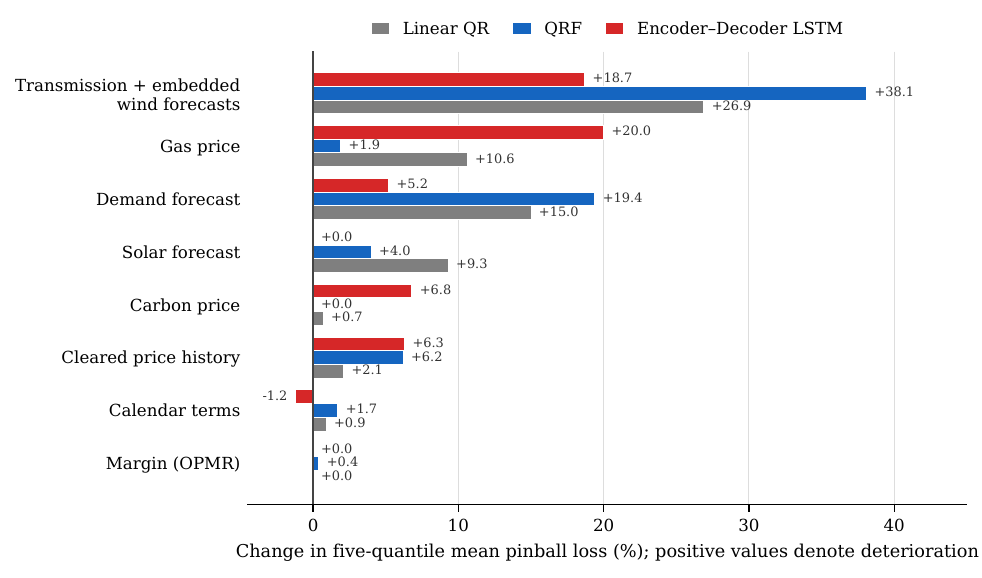}
  \caption[Grouped information-family ablation over the 2025 test
  period.]{Grouped information-family ablation over the 2025 test period. Bars
  show the percentage change in five-quantile mean pinball loss after removing
  one predictor family and refitting under the same rolling protocol; positive
  values denote deterioration. LSTM values average within-seed changes across
  seeds 42, 7 and 101.}
  \label{fig:ablation}
\end{figure}

Figure~\ref{fig:ablation} reports the grouped ablation. Removing the
transmission and embedded wind forecasts produced the largest deterioration for
both tabular models, at 26.9\% for linear quantile regression and 38.1\% for
QRF, and a substantial but smaller deterioration of 18.7\% for the LSTM. The
demand forecast follows the same ordering, at 15.0\%, 19.4\% and 5.2\%
respectively.

The gas-price channel reverses it. Removing gas prices deteriorated LSTM
performance by 20.0\%, its largest single effect, compared with 10.6\% for
linear quantile regression and 1.9\% for QRF. Carbon prices show the same
pattern more weakly, at 6.8\% for the LSTM against approximately zero for both
tabular models. Cleared price history mattered similarly to QRF and the LSTM, at
6.2\% and 6.3\%, while solar forecasts, calendar terms and the margin report
produced effects close to zero for all three models.

The two model classes therefore do not rest on the same information. Both depend
heavily on the wind forecasts, but the LSTM extracted substantially more from
the commodity channels while QRF extracted more from the demand forecast and the
cleared price history. This difference in information base accompanies the
difference in conditional performance documented in
Section~\ref{sec:res_conditional}.

\FloatBarrier

%==============================================================================
\section{Discussion}
\label{sec:discussion}

\subsection{Opposite failure modes in the two model classes}

The central empirical result is that two models with statistically
indistinguishable aggregate skill behaved very differently in the price regimes
that motivate probabilistic forecasting. The construction of each model offers a
plausible explanation. A quantile regression forest derives every quantile from
a weighted empirical distribution of training responses
\citep{meinshausen2006qrf}, so its upper quantiles are drawn from prices that
have already been observed within the rolling window. Spikes that recur within
that window are therefore represented directly in the forecast distribution,
while prices beyond the observed support cannot be produced at all. The
Encoder–Decoder LSTM estimates each quantile by minimising pinball loss over a
728-day window in which hours above 200 GBP/MWh are rare, and the resulting
upper quantile function is smooth in the inputs. It generalises well where
observations are dense, which is the centre of the distribution, and reverts
towards typical levels where they are sparse.

The ablation results are consistent with this account and add a second
dimension to it. QRF depended most on the wind and demand forecasts, the
variables that determine the supply-demand balance, and comparatively little on
gas prices. The LSTM depended most on the gas-price channel, which is
consistent with gas frequently setting the marginal price in GB
\citep{ofgem2026wholesale,zakeri2023gas} while wind determines how much demand
must be met by thermal generation \citep{sensfuss2008meritorder}. The forest
appears to locate spike conditions by recognising tight-margin states
resembling those in its training leaves, whereas the network appears to track
the cost level around which prices are formed. These are complementary rather
than competing accounts of price formation, and neither model captures both.

\subsection{Recalibration and combination are not substitutes}

Sections~\ref{sec:res_conformal} and~\ref{sec:res_qra} compare the two available
remedies on the same forecasts. Rolling conformal recalibration achieved what it
targets, moving 90\% coverage from 0.895 to 0.900 and balancing the tail miss
rates at 4.8\% and 5.2\%, and it was clearly useful at the 80\% level where a
genuine coverage error existed, raising 0.778 to 0.801 with less widening than
the static variant. That result is consistent with the adaptive-calibration
benefit reported by \citet{brusaferri2025conformal} and with the coverage range
reported by \citet{kath2021conformal}. It did not, however, improve the Winkler
score at either level, which is the expected outcome when an already reliable
interval is widened.

The limitation is structural rather than a matter of tuning. The conditional
checks show that static conformal prediction did not change high-price coverage,
while rolling calibration raised coverage above 200 GBP/MWh only from 0.12 to
0.19. Conformal corrections are quantiles of pooled conformity errors and are applied uniformly
across delivery hours, so they can move the bounds but cannot make the interval
wide in the hours that will spike and narrow in the hours that will not. QRA
operates on different material. By estimating quantile-level weights on matched
out-of-sample forecasts, it inherited QRF's spike information in the tails and
the LSTM's central accuracy, and the fitted weights reproduced that division
without being instructed to. LEAR-QRA then supplied a smaller contribution in
the centre, reducing CRPS by a further 1.8\% relative to the two-model
combination. Coverage above 200 GBP/MWh rose from 0.12 to 0.65 while MPIL changed
only from 32.5 to 32.8 GBP/MWh and marginal coverage reached 0.900, so the gain
came from better placement of interval width rather than from more of it. The
Winkler score fell from 51.2 to 44.1.

These findings support earlier evidence for forecast combination from the
GEFCom2014 price track \citep{hong2016gefcom} and from QRA improvements over a
distributional neural model \citep{lipiecki2024postprocessing}. Unlike the
under-coverage reported by \citet{cornell2024nem} and
\citet{oconnor2025conformal}, coverage here remained near nominal, plausibly
because the constituents were already well calibrated and QRA was re-estimated
weekly on matched out-of-sample data.

\subsection{Comparison with published GB evidence}

The Encoder–Decoder LSTM compares favourably with the recent GB literature. Its MAE of
7.89 GBP/MWh is 44.7\% below the LSTM MAE reported by \citet{cooke2025ukepf},
and its CRPS of 5.59 GBP/MWh is 66.7\% below the EPEX CRPS reported by
\citet{agakishiev2025multivariate}. In contrast to the latter study, the LSTM
was the strongest individual model here, which suggests that architecture, input
alignment and evaluation design affect whether recurrent models outperform
simpler alternatives. These comparisons are contextual rather than controlled:
the studies differ in period, resolution, price target and information
boundary, and 2025 was not a year of comparable market stress to some earlier
test windows. They establish that the models evaluated here are not weak
implementations, not that they are superior to the published alternatives.
The standard LEAR-QRA benchmark was also weaker than the LSTM and QRF here, with
CRPS of 6.52 GBP/MWh, but improved the wider combination significantly. This
distinguishes constituent diversity from stand-alone accuracy: a weaker model
can add value when its forecast errors are not fully redundant.

\subsection{Implications for practice}

For a forecast user whose exposure is concentrated in extreme hours, the
practical implication is that model selection on CRPS or marginal coverage is
insufficient and can be actively misleading. In this evaluation, selecting on
CRPS alone would have chosen the model that covered 12\% of the most extreme
outcomes over one that covered 71\%. Conditional coverage at thresholds fixed in
advance is inexpensive to compute from forecasts that already exist, and it
should be reported alongside aggregate scores rather than as a supplementary
diagnostic.

The second implication concerns what to do about an unreliable interval. Where
the failure is a level error in marginal coverage, conformal recalibration is an
appropriate and low-cost remedy, and the rolling variant is preferable under
temporal change. Where the failure is conditional, and specifically where the
interval fails in identifiable market states, recalibration will not resolve it
and a combination that includes a model with different structural properties is
the more promising route.

The accuracy gain from that combination can also be placed on a simple monetary
scale. For a fixed-volume position, multiplying the median price forecast by
traded volume gives a pre-auction estimate of its cost or revenue. From the
median MAEs in Table~\ref{tab:headline}, three-model QRA reduced the average
absolute price error by 0.63~GBP/MWh relative to the Encoder--Decoder LSTM (7.89
to 7.26) and by 1.12~GBP/MWh relative to QRF (8.38 to 7.26). For an illustrative
10~MWh position in one delivery hour, this corresponds to reductions of
6.30~GBP and 11.20~GBP in the average absolute error of the estimated position
value, respectively. The conditional-coverage results also show that three-model
QRA retained approximately 90\% of the coverage improvement that QRF provided
over the LSTM above 200~GBP/MWh. It therefore combined more accurate year-round
valuation with most of QRF's upper-tail advantage. These monetary amounts
represent reductions in pre-auction valuation error rather than realised trading
profit, since the study does not model the positions taken from the forecasts.

\subsection{Limitations}

The study covers one market and one complete test year, so the model rankings
cannot be assumed to generalise beyond 2025 GB prices. The conditional coverage
results above 200 GBP/MWh rest on 52 hours, and although the difference between
0.12 and 0.71 is large, the estimates are imprecise and are not yet accompanied
by confidence intervals or a significance test at the tail quantiles. CRPS was
approximated from eleven quantiles rather than a denser grid. The LEAR-QRA
benchmark could not reproduce the full range of calibration windows used by
\citet{lago2021benchmark} on longer benchmark datasets: very short windows made
the AIC-selected high-dimensional regressions numerically unstable near the
predictor count, while the present sample could not support their longest
multi-year windows. QRA used linear weights pooled
across delivery hours, and its results are predictive rather than causal. The hourly forecasts represent marginal distributions rather than joint temporal scenarios, so while the results demonstrate marked improvements in tail coverage, they do not directly quantify end-to-end financial performance across multi-settlement trading or portfolio risk management workflows. Finally, lagged
settlement outturns use a later revision vintage than would have been available
at the gate, as noted in Section~\ref{sec:data}.

%==============================================================================
\section{Conclusions}
\label{sec:conclusion}

This study evaluated probabilistic forecasts of hourly EPEX GB day-ahead prices
over a full held-out calendar year, applying a common pre-auction information
boundary and rolling-origin design to a historical benchmark, linear quantile
regression, LEAR-QRA, a quantile regression forest and an Encoder-Decoder
quantile LSTM, and comparing two routes to more reliable intervals.

The principal finding is that aggregate probabilistic scores were uninformative
about tail reliability, which is what matters most in this market. The LSTM and
QRF differed by 4.4\% in CRPS, a difference that was not statistically
significant, and both produced reliability diagrams close to the diagonal.
Above 200 GBP/MWh the LSTM covered 0.12 of outcomes and QRF 0.71. The divergence
is consistent with the construction of the two model classes, since the forest
derives its quantiles from realised prices within its rolling window while the
network learns a smooth upper tail from few spike observations, and it is
reflected in the different predictor families on which the two models depend.

The pattern was not confined to the recurrent model. LEAR-QRA attained marginal
coverage of 0.893, the closest to nominal of any model assessed, and covered
0.06 of the same hours. A forecast can therefore satisfy the calibration
diagnostic most commonly reported while failing almost completely in the regime
that motivates probabilistic forecasting.

The two remedies for unreliable intervals addressed different failures. Rolling
conformal recalibration brought marginal coverage to nominal and balanced the
tail miss rates, and was clearly useful at the under-covering 80\% level, but it
did not improve the Winkler score at either level and cannot by construction
supply information about which hours will spike.

Combining the quantile regression forest and the LSTM by Quantile Regression
Averaging raised coverage above 200 GBP/MWh from 0.12 to 0.65 while interval
width changed only from 32.5 to 32.8 GBP/MWh, and reduced the Winkler score from
51.2 for the LSTM to 44.1. Measured against the LEAR benchmark on identical
data, the same combination reduced CRPS by 19.8\% and the Winkler score by
38.1\%. Adding LEAR-QRA as a third constituent lowered CRPS further to 5.14
GBP/MWh, a statistically significant gain of 1.8\%, without altering the tail
results. LEAR-QRA was the weakest constituent on every metric reported, so the
value of a constituent derives from the independence of its errors rather than
from its stand-alone accuracy. Recalibration and combination are therefore
complements rather than alternatives, and the choice between them should follow
from whether the observed failure is marginal or conditional.

Two directions follow. Evaluations of probabilistic price forecasts should
report conditional coverage at thresholds fixed in advance alongside aggregate
scores, since the additional cost is negligible and the additional information
is decisive. Beyond that, the complementarity documented here should be
evaluated in downstream decision frameworks, including risk-constrained
scheduling, cash-flow-at-risk modelling and trading backtests. Such work would
quantify how improved tail reliability translates into avoided tail losses and
operational value.

%==============================================================================
\section*{CRediT authorship contribution statement}

Anirudh Agrawal: Conceptualization, Methodology, Software, Formal analysis, Data curation, Visualization, Writing (original draft). Aidan O'Sullivan: Conceptualization, Methodology,
Supervision, Writing (review and editing).

\section*{Generative AI Statement}
Open AI Codex and Anthropic Claude assisted with code development and verification. The lead author reviewed and validated all retained code and results.

\section*{Declaration of competing interest}
Habitat Energy, a battery storage optimisation company active in the GB wholesale electricity market, supported this work by providing access to market data and by commenting on the study during its development. Habitat Energy had no role in the design of the study, in the analysis or interpretation of the
results, in the writing of the manuscript, or in the decision to submit it for publication. The authors received no funding, payment or other financial benefit from Habitat Energy in connection with this work, and declare no other competing financial interests or personal relationships that could have
appeared to influence the work reported in this paper.

\section*{Data availability}
EPEX SPOT day-ahead auction prices are licensed and cannot be redistributed by
the authors. All system inputs are publicly available: demand and transmission
wind forecast vintages, balancing and generation outturns and REMIT records from
Elexon BMRS \citep{elexon2026insights}; embedded renewable forecasts,
interconnector outturns and Operational Planning Margin Reports from NESO
\citep{neso2026opmr,neso2026dataportal}; and daily gas and carbon futures proxies
from Investing.com \citep{investing2026ukgas,investing2026ukcarbon}.
The Elexon and NESO sources were last accessed on 9 August 2026, and the gas and
carbon series on 24 July 2026. The modelling and evaluation code is archived at
\url{https://doi.org/10.5281/zenodo.22836687}.

\section*{Acknowledgements}
The authors would like to thank Duncan Fraser of Habitat Energy for feedback and supporting this project with data access. 
%==============================================================================
\appendix

\section{Supplementary results}
\label{app:supplementary}

\begin{table}[htbp]
\centering
\footnotesize
\setlength{\tabcolsep}{4pt}
\renewcommand{\arraystretch}{1.1}
\caption[Probabilistic performance at the 80\% interval during 2025.]{
Probabilistic performance at the 80\% interval during 2025, the counterpart to
Table~\ref{tab:headline}. CRPS and MAE are unchanged and are repeated for
reference.}
\label{tab:headline_80}
\begin{tabular*}{\textwidth}{@{\extracolsep{\fill}}lrrrrrrr@{}}
\toprule
Model & CRPS & MAE & Cov. & MPIL & Winkler & Cov. & Cov. \\
      & (GBP/MWh) & (GBP/MWh) & at 80\% & (GBP/MWh) & (GBP/MWh) & $<0$ & $>150$ \\
\midrule
Historical benchmark & 15.55 & 20.89 & 0.766 & 65.5 & 116.2 & 0.00 & 0.00 \\
Linear QR            & 6.89  & 9.68  & 0.704 & 25.8 & 49.6  & 0.59 & 0.29 \\
LEAR-QRA             & 6.52  & 8.72  & 0.788 & 23.9 & 51.3  & 0.32 & 0.22 \\
QRF                  & 5.85  & 8.38  & 0.831 & 27.9 & 38.9  & 0.63 & 0.63 \\
Encoder–Decoder LSTM       & 5.59  & 7.89  & 0.778 & 23.0 & 39.5  & 0.73 & 0.42 \\
Two-model QRA        & 5.23  & 7.45  & 0.797 & 24.8 & 35.5  & 0.74 & 0.70 \\
Three-model QRA      & 5.14  & 7.26  & 0.804 & 24.5 & 35.4  & 0.73 & 0.67 \\
\bottomrule
\end{tabular*}
\end{table}

\begin{table}[htbp]
\centering
\footnotesize
\setlength{\tabcolsep}{6pt}
\renewcommand{\arraystretch}{1.1}
\caption[Conditional 90\% interval coverage above 150 and 200 GBP/MWh.]{
Conditional 90\% interval coverage above 150 and 200 GBP/MWh during 2025, with
the number of qualifying delivery hours.}
\label{tab:conditional_tail}
\begin{tabular}{lrr}
\toprule
Model & Cov. $>150$ ($n=123$) & Cov. $>200$ ($n=52$) \\
\midrule
Historical benchmark & 0.15 & 0.00 \\
Linear QR & 0.44 & 0.08 \\
LEAR-QRA & 0.28 & 0.06 \\
QRF & 0.76 & 0.71 \\
Encoder–Decoder LSTM & 0.51 & 0.12 \\
Static CQR & 0.51 & 0.12 \\
Rolling CQR & 0.55 & 0.19 \\
Two-model QRA & 0.79 & 0.65 \\
Three-model QRA & 0.79 & 0.65 \\
\bottomrule
\end{tabular}
\end{table}

\begin{table}[htbp]
\centering
\footnotesize
\setlength{\tabcolsep}{5pt}
\renewcommand{\arraystretch}{1.1}
\caption{LEAR-QRA window and three-model QRA regularisation sensitivities over
the 2025 held-out year.}
\label{tab:qra_sensitivities}
\begin{tabular}{lrrrrr}
\toprule
Specification & CRPS & MAE & Cov. 90\% & MPIL 90\% & Winkler 90\% \\
\midrule
LEAR-QRA, 364/546/728 days & 6.522 & 8.715 & 0.893 & 37.4 & 71.2 \\
LEAR-QRA, plus 273 days & 6.529 & 8.724 & 0.892 & 37.3 & 71.5 \\
Two-model QRA & 5.234 & 7.454 & 0.893 & 32.8 & 44.1 \\
Three-model QRA & 5.138 & 7.256 & 0.900 & 32.8 & 44.1 \\
Regularised three-model QRA & 5.138 & 7.256 & 0.900 & 32.8 & 44.1 \\
\bottomrule
\end{tabular}
\end{table}

\begin{table}[htbp]
\centering
\small
\renewcommand{\arraystretch}{1.12}
\caption{Encoder–Decoder LSTM training settings.}
\label{tab:lstm_training}
\begin{tabular}{p{0.38\linewidth}p{0.52\linewidth}}
\toprule
Setting & Specification \\
\midrule
Recurrent units & 64 in the encoder and 64 in the decoder \\
Activations & LSTM $\tanh$ activation with sigmoid recurrent gates; linear output \\
Dropout & 0.1 after the decoder \citep{srivastava2014dropout} \\
Loss function & Pinball loss; one network for each quantile \\
Optimiser and learning rate & Adam; $10^{-3}$ \citep{kingma2015adam} \\
Batch size & 16 \\
Maximum epochs & 200 \\
Early stopping & Patience of 10 epochs; best weights restored \\
Within-window validation & Final 14 days of each rolling estimation window \\
Random seeds & 42, 7 and 101; forecasts averaged \\
\bottomrule
\end{tabular}
\end{table}
\FloatBarrier

\begin{figure}[htbp]
  \centering
  \begin{subfigure}{0.32\textwidth}
    \centering\includegraphics[width=\textwidth]{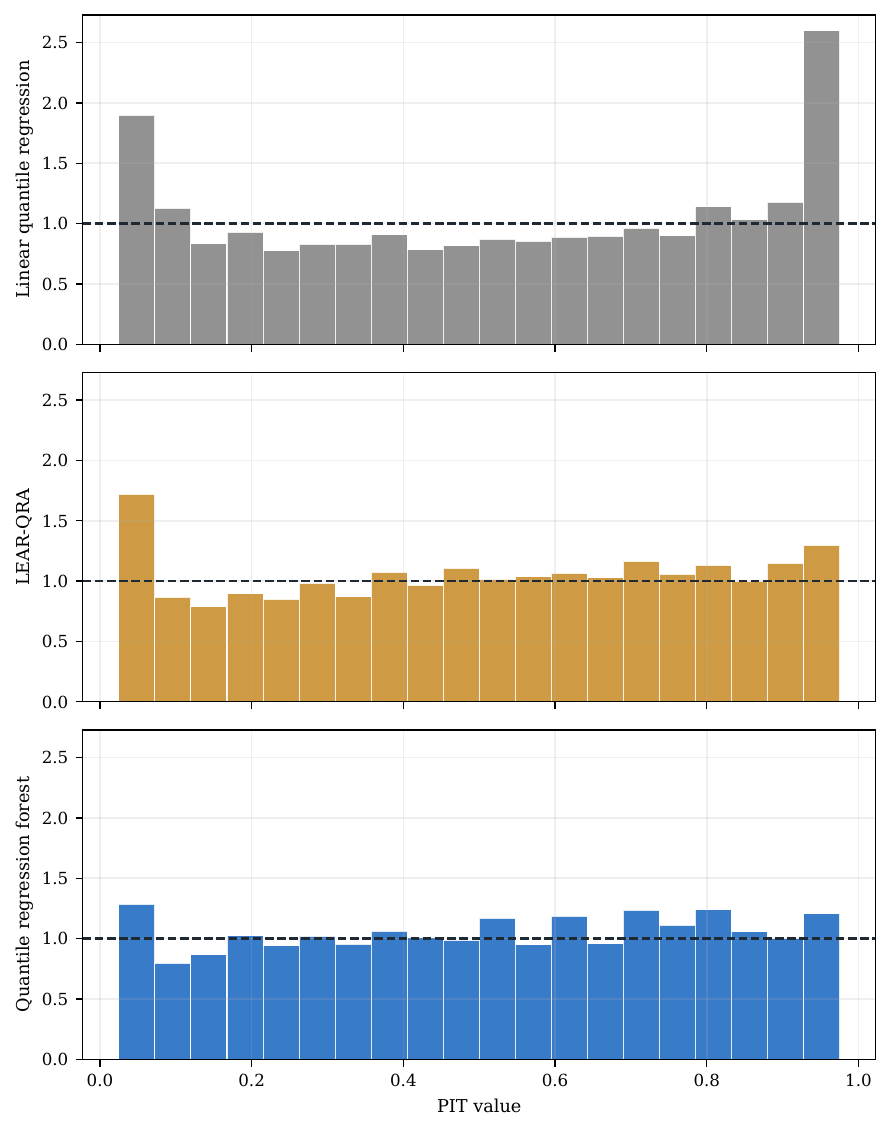}
    \caption{Linear QR, LEAR-QRA and QRF.}
  \end{subfigure}\hfill
  \begin{subfigure}{0.32\textwidth}
    \centering\includegraphics[width=\textwidth]{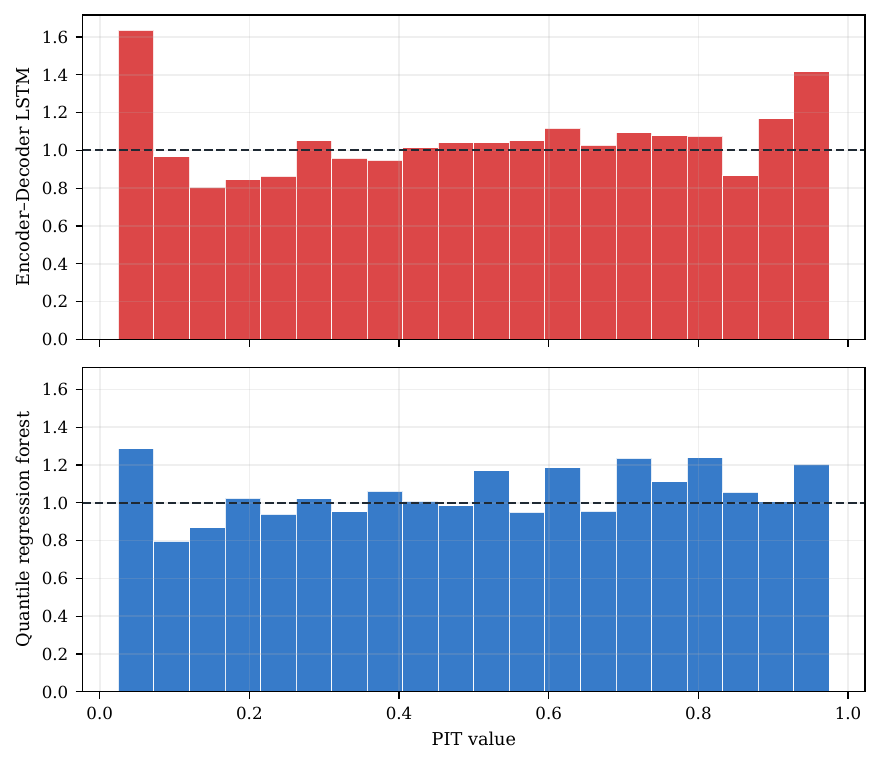}
    \caption{LSTM and QRF.}
  \end{subfigure}\hfill
  \begin{subfigure}{0.32\textwidth}
    \centering\includegraphics[width=\textwidth]{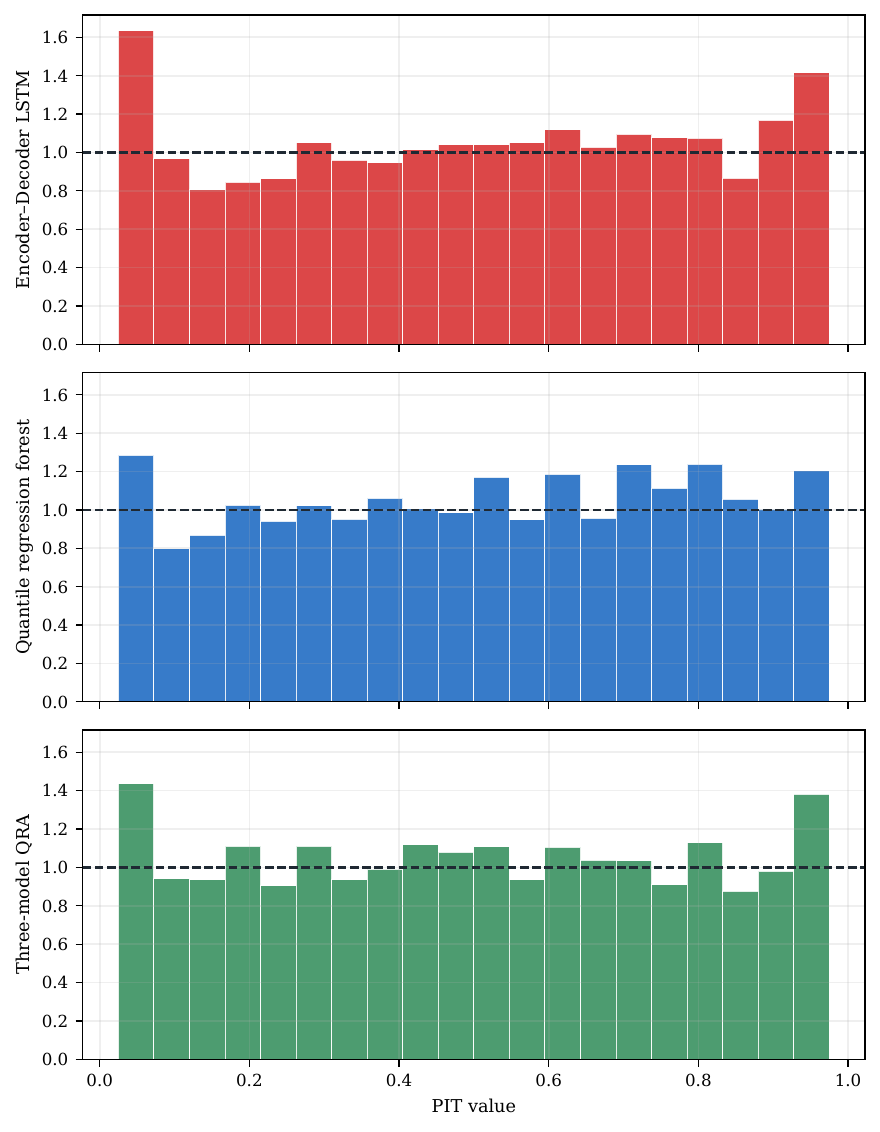}
    \caption{Three-model QRA and its principal constituents.}
  \end{subfigure}
  \caption[Approximate PIT histograms during 2025.]{Approximate PIT histograms
  during 2025. The dashed line denotes a uniform reference density.}
  \label{fig:appendix_pit}
\end{figure}

\begin{figure}[htbp]
  \centering
  \includegraphics[width=\textwidth]{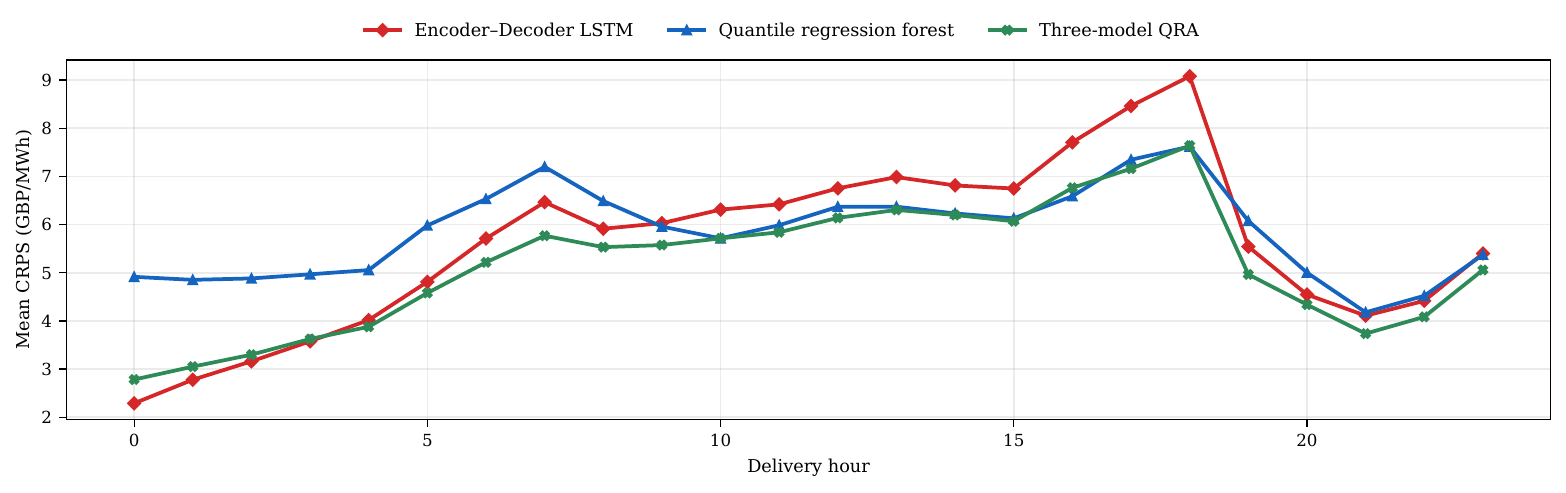}
  \caption{Mean CRPS by delivery hour for three-model QRA and its principal constituent models during
  2025.}
  \label{fig:appendix_rq4_hourly_crps}
\end{figure}

% Generated by 05_code/scripts/make_feature_exhibits.py -- do not hand-edit.
\begin{landscape}
\footnotesize
\setlength{\tabcolsep}{5pt}
\begin{longtable}{p{0.25\linewidth}p{0.43\linewidth}cccc}
\caption[Full predictor membership and forecast-time availability by model.]{Full predictor membership and forecast-time availability by model. A check mark
denotes a direct input and a dash denotes exclusion. The encoder check mark on
\texttt{price\_lag1d} represents the complete 24-hour D$-$1 cleared-price
profile rather than one scalar input. The per-hour linear regressions omit the
three within-day hour terms because they are constant in each hourly model.}
\label{tab:feature_membership_full}\\
\toprule
Feature or channel & Construction and availability & Linear QR & QRF & Encoder & Decoder \\
\midrule
\endfirsthead
\multicolumn{6}{l}{\tablename\ \thetable\ continued}\\
\toprule
Feature or channel & Construction and availability & Linear QR & QRF & Encoder & Decoder \\
\midrule
\endhead
\midrule
\multicolumn{6}{l}{\itshape calendar}\\*
dayofweek & Calendar known before gate closure & \checkmark & \checkmark & -- & \checkmark\\
hour & Calendar known before gate closure & -- & \checkmark & -- & --\\
hour\_cos & Calendar known before gate closure & -- & \checkmark & -- & \checkmark\\
hour\_sin & Calendar known before gate closure & -- & \checkmark & -- & \checkmark\\
is\_bank\_holiday & Calendar known before gate closure & \checkmark & \checkmark & -- & \checkmark\\
month\_cos & Calendar known before gate closure & \checkmark & \checkmark & -- & \checkmark\\
month\_sin & Calendar known before gate closure & \checkmark & \checkmark & -- & \checkmark\\
\midrule
\multicolumn{6}{l}{\itshape forecast fundamentals}\\*
demand\_forecast & Latest D forecast available by the D$-$1 gate & \checkmark & \checkmark & -- & \checkmark\\
embedded\_solar\_forecast & Latest D forecast available by the D$-$1 gate & \checkmark & \checkmark & -- & \checkmark\\
embedded\_wind\_forecast & Latest D forecast available by the D$-$1 gate & \checkmark & \checkmark & -- & \checkmark\\
low\_residual\_load\_flag & Derived from eligible delivery-D demand and renewable forecasts & \checkmark & \checkmark & -- & --\\
low\_residual\_load\_mw & Derived from eligible delivery-D demand and renewable forecasts & \checkmark & \checkmark & -- & --\\
net\_demand\_forecast & Derived from eligible delivery-D demand and renewable forecasts & \checkmark & \checkmark & -- & --\\
renewable\_share\_forecast & Derived from eligible delivery-D demand and renewable forecasts & \checkmark & \checkmark & -- & --\\
transmission\_wind\_forecast & Latest D forecast available by the D$-$1 gate & \checkmark & \checkmark & -- & \checkmark\\
\midrule
\multicolumn{6}{l}{\itshape commodities}\\*
carbon\_price & Last eligible close, no later than D$-$2 & \checkmark & \checkmark & -- & \checkmark\\
gas\_price & Last eligible close, no later than D$-$2 & \checkmark & \checkmark & -- & \checkmark\\
\midrule
\multicolumn{6}{l}{\itshape OPMR margin}\\*
opmr\_national\_surplus & Latest valid OPMR vintage, no later than D$-$2 & \checkmark & \checkmark & -- & --\\*
opmr\_surplus\_norm & Latest valid OPMR vintage, no later than D$-$2 & \checkmark & \checkmark & -- & \checkmark\\*
opmr\_surplus\_rev\_7d & Latest valid OPMR vintage, no later than D$-$2 & \checkmark & \checkmark & -- & --\\
\midrule
\multicolumn{6}{l}{\itshape REMIT news}\\*
remit\_events\_published\_1d & Events published by the D$-$1 gate; trailing 1-day count & \checkmark & \checkmark & -- & --\\
remit\_events\_published\_7d & Events published by the D$-$1 gate; trailing 7-day count & \checkmark & \checkmark & -- & --\\
\midrule
\multicolumn{6}{l}{\itshape cleared auction-price history}\\*
price\_lag1d & Cleared auction price for D$-$1 & \checkmark & \checkmark & \checkmark & --\\
price\_lag2d & Cleared auction price for D$-$2 & \checkmark & \checkmark & -- & --\\
price\_lag7d & Cleared auction price for D$-$7 & \checkmark & \checkmark & -- & --\\
\midrule
\multicolumn{6}{l}{\itshape balancing state}\\*
netImbalanceVolume\_lag2d & Published settlement outturn, lagged D$-$2 & \checkmark & \checkmark & -- & --\\
netImbalanceVolume\_lag7d & Published settlement outturn, lagged D$-$7 & \checkmark & \checkmark & -- & --\\
systemSellPrice\_lag2d & Published settlement outturn, lagged D$-$2 & \checkmark & \checkmark & -- & --\\
systemSellPrice\_lag7d & Published settlement outturn, lagged D$-$7 & \checkmark & \checkmark & -- & --\\
\midrule
\multicolumn{6}{l}{\itshape generation mix}\\*
gen\_ccgt\_mw\_lag2d & Published generation outturn, lagged D$-$2 & \checkmark & \checkmark & -- & --\\
gen\_nuclear\_mw\_lag2d & Published generation outturn, lagged D$-$2 & \checkmark & \checkmark & -- & --\\
gen\_wind\_mw\_lag2d & Published generation outturn, lagged D$-$2 & \checkmark & \checkmark & -- & --\\
\midrule
\multicolumn{6}{l}{\itshape interconnectors}\\*
net\_interconnector\_flow\_lag2d & Published flow outturn, lagged D$-$2 & \checkmark & \checkmark & -- & --\\
net\_interconnector\_flow\_lag7d & Published flow outturn, lagged D$-$7 & \checkmark & \checkmark & -- & --\\
\bottomrule
\end{longtable}
\end{landscape}

\FloatBarrier

%==============================================================================
\bibliographystyle{elsarticle-num-names}
\bibliography{references}

@article{lago2021benchmark,
  author  = {Lago, Jesus and Marcjasz, Grzegorz and De Schutter, Bart and Weron, Rafa{\l}},
  title   = {Forecasting day-ahead electricity prices: A review of state-of-the-art algorithms, best practices and an open-access benchmark},
  journal = {Applied Energy},
  year    = {2021},
  volume  = {293},
  pages   = {116983},
  doi     = {10.1016/j.apenergy.2021.116983}
}

@article{nowotarski2018review,
  author  = {Nowotarski, Jakub and Weron, Rafa{\l}},
  title   = {Recent advances in electricity price forecasting: A review of probabilistic forecasting},
  journal = {Renewable and Sustainable Energy Reviews},
  year    = {2018},
  volume  = {81},
  number  = {Part 1},
  pages   = {1548--1568},
  doi     = {10.1016/j.rser.2017.05.234}
}

@article{nowotarski2015qra,
  author  = {Nowotarski, Jakub and Weron, Rafa{\l}},
  title   = {Computing electricity spot price prediction intervals using quantile regression and forecast averaging},
  journal = {Computational Statistics},
  year    = {2015},
  volume  = {30},
  number  = {3},
  pages   = {791--803},
  doi     = {10.1007/s00180-014-0523-0}
}

@article{uniejewski2021lqra,
  author  = {Uniejewski, Bartosz and Weron, Rafa{\l}},
  title   = {Regularized quantile regression averaging for probabilistic electricity price forecasting},
  journal = {Energy Economics},
  year    = {2021},
  volume  = {95},
  pages   = {105121},
  doi     = {10.1016/j.eneco.2021.105121}
}

@article{maciejowska2016fqra,
  author  = {Maciejowska, Katarzyna and Nowotarski, Jakub and Weron, Rafa{\l}},
  title   = {Probabilistic forecasting of electricity spot prices using factor quantile regression averaging},
  journal = {International Journal of Forecasting},
  year    = {2016},
  volume  = {32},
  number  = {3},
  pages   = {957--965},
  doi     = {10.1016/j.ijforecast.2014.12.004}
}

@article{agakishiev2025multivariate,
  author  = {Agakishiev, Ilyas and H{\"a}rdle, Wolfgang Karl and Kopa, Milos and Kozmik, Karel and Petukhina, Alla},
  title   = {Multivariate probabilistic forecasting of electricity prices with trading applications},
  journal = {Energy Economics},
  year    = {2025},
  volume  = {141},
  pages   = {108008},
  doi     = {10.1016/j.eneco.2024.108008}
}

@mastersthesis{cooke2025ukepf,
  author  = {Cooke, Stephen},
  title   = {Enhancing {UK} Electricity Price Forecasting Using Deep Learning},
  school  = {CCT College Dublin},
  year    = {2025},
  type    = {{MSc} thesis},
  doi     = {10.63227/652.299.97},
  url     = {https://arc.cct.ie/ict/135/},
  accessed = {22 July 2026}
}

@article{ugurlu2018rnn,
  author  = {Ugurlu, Umut and Oksuz, Ilkay and Tas, Oktay},
  title   = {Electricity Price Forecasting Using Recurrent Neural Networks},
  journal = {Energies},
  year    = {2018},
  volume  = {11},
  number  = {5},
  pages   = {1255},
  doi     = {10.3390/en11051255}
}

@article{lehna2022comparison,
  author  = {Lehna, Malte and Scheller, Fabian and Herwartz, Helmut},
  title   = {Forecasting day-ahead electricity prices: A comparison of time series and neural network models taking external regressors into account},
  journal = {Energy Economics},
  year    = {2022},
  volume  = {106},
  pages   = {105742},
  doi     = {10.1016/j.eneco.2021.105742}
}

@article{uniejewski2016varsel,
  author  = {Uniejewski, Bartosz and Nowotarski, Jakub and Weron, Rafa{\l}},
  title   = {Automated variable selection and shrinkage for day-ahead electricity price forecasting},
  journal = {Energies},
  year    = {2016},
  volume  = {9},
  number  = {8},
  pages   = {621},
  doi     = {10.3390/en9080621}
}

@article{cornell2024nem,
  author  = {Cornell, Cameron and Dinh, Nam Trong and Pourmousavi, S. Ali},
  title   = {A probabilistic forecast methodology for volatile electricity prices in the {Australian} {National} {Electricity} {Market}},
  journal = {International Journal of Forecasting},
  year    = {2024},
  volume  = {40},
  number  = {4},
  pages   = {1421--1437},
  doi     = {10.1016/j.ijforecast.2023.12.003}
}

@article{gneiting2007jrssb,
  author  = {Gneiting, Tilmann and Balabdaoui, Fadoua and Raftery, Adrian E.},
  title   = {Probabilistic forecasts, calibration and sharpness},
  journal = {Journal of the Royal Statistical Society: Series B},
  year    = {2007},
  volume  = {69},
  number  = {2},
  pages   = {243--268},
  doi     = {10.1111/j.1467-9868.2007.00587.x}
}

@article{gneiting2007jasa,
  author  = {Gneiting, Tilmann and Raftery, Adrian E.},
  title   = {Strictly proper scoring rules, prediction, and estimation},
  journal = {Journal of the American Statistical Association},
  year    = {2007},
  volume  = {102},
  number  = {477},
  pages   = {359--378},
  doi     = {10.1198/016214506000001437}
}

@article{matheson1976crps,
  author  = {Matheson, James E. and Winkler, Robert L.},
  title   = {Scoring rules for continuous probability distributions},
  journal = {Management Science},
  year    = {1976},
  volume  = {22},
  number  = {10},
  pages   = {1087--1096},
  doi     = {10.1287/mnsc.22.10.1087}
}

@article{diebold1995comparing,
  author  = {Diebold, Francis X. and Mariano, Roberto S.},
  title   = {Comparing predictive accuracy},
  journal = {Journal of Business \& Economic Statistics},
  year    = {1995},
  volume  = {13},
  number  = {3},
  pages   = {253--263},
  doi     = {10.1080/07350015.1995.10524599}
}

@article{neweywest1987hac,
  author  = {Newey, Whitney K. and West, Kenneth D.},
  title   = {A Simple, Positive Semi-Definite, Heteroskedasticity and Autocorrelation Consistent Covariance Matrix},
  journal = {Econometrica},
  year    = {1987},
  volume  = {55},
  number  = {3},
  pages   = {703--708},
  doi     = {10.2307/1913610}
}

@article{chernozhukov2010rearrangement,
  author  = {Chernozhukov, Victor and Fern{\'a}ndez-Val, Iv{\'a}n and Galichon, Alfred},
  title   = {Quantile and Probability Curves Without Crossing},
  journal = {Econometrica},
  year    = {2010},
  volume  = {78},
  number  = {3},
  pages   = {1093--1125},
  doi     = {10.3982/ECTA7880}
}

@article{diebold1998pit,
  author  = {Diebold, Francis X. and Gunther, Todd A. and Tay, Anthony S.},
  title   = {Evaluating density forecasts with applications to financial risk management},
  journal = {International Economic Review},
  year    = {1998},
  volume  = {39},
  number  = {4},
  pages   = {863--883},
  doi     = {10.2307/2527342}
}

@article{pinson2007windeval,
  author  = {Pinson, Pierre and Nielsen, Henrik Aa. and M{\o}ller, Jan K. and Madsen, Henrik and Kariniotakis, George N.},
  title   = {Non-parametric probabilistic forecasts of wind power: required properties and evaluation},
  journal = {Wind Energy},
  year    = {2007},
  volume  = {10},
  number  = {6},
  pages   = {497--516},
  doi     = {10.1002/we.230}
}

@article{hamill2001rankhist,
  author  = {Hamill, Thomas M.},
  title   = {Interpretation of rank histograms for verifying ensemble forecasts},
  journal = {Monthly Weather Review},
  year    = {2001},
  volume  = {129},
  number  = {3},
  pages   = {550--560},
  doi     = {10.1175/1520-0493(2001)129<0550:IORHFV>2.0.CO;2}
}

@article{winkler1972interval,
  author  = {Winkler, Robert L.},
  title   = {A decision-theoretic approach to interval estimation},
  journal = {Journal of the American Statistical Association},
  year    = {1972},
  volume  = {67},
  number  = {337},
  pages   = {187--191},
  doi     = {10.1080/01621459.1972.10481224}
}

@article{lerch2017dilemma,
  author  = {Lerch, Sebastian and Thorarinsdottir, Thordis L. and Ravazzolo, Francesco and Gneiting, Tilmann},
  title   = {Forecaster's dilemma: extreme events and forecast evaluation},
  journal = {Statistical Science},
  year    = {2017},
  volume  = {32},
  number  = {1},
  pages   = {106--127},
  doi     = {10.1214/16-STS588}
}

@article{bjerregaard2021mvreview,
  author  = {Bjerreg{\aa}rd, Mathias B. and M{\o}ller, Jan K. and Madsen, Henrik},
  title   = {An introduction to multivariate probabilistic forecast evaluation},
  journal = {Energy and AI},
  year    = {2021},
  volume  = {4},
  pages   = {100058},
  doi     = {10.1016/j.egyai.2021.100058}
}

@article{messner2020windeval,
  author  = {Messner, Jakob W. and Pinson, Pierre and Browell, Jethro and Bjerreg{\aa}rd, Mathias B. and Schicker, Irene},
  title   = {Evaluation of wind power forecasts--An up-to-date view},
  journal = {Wind Energy},
  year    = {2020},
  volume  = {23},
  number  = {6},
  pages   = {1461--1481},
  doi     = {10.1002/we.2497}
}

@misc{allen2025conditionalinterval,
  url     = {https://arxiv.org/abs/2508.18034},
  accessed = {10 July 2026},
  author       = {Allen, Sam and Burnello, Julia and Ziegel, Johanna},
  title        = {Assessing the conditional calibration of interval forecasts using decompositions of the interval score},
  year         = {2025},
  eprint       = {2508.18034},
  archivePrefix = {arXiv},
  primaryClass = {stat.ME}
}

@article{sensfuss2008meritorder,
  author  = {Sensfu{\ss}, Frank and Ragwitz, Mario and Genoese, Massimo},
  title   = {The merit-order effect: A detailed analysis of the price effect of renewable electricity generation on spot market prices in {Germany}},
  journal = {Energy Policy},
  year    = {2008},
  volume  = {36},
  number  = {8},
  pages   = {3086--3094},
  doi     = {10.1016/j.enpol.2008.03.035}
}

@article{hong2016gefcom,
  author  = {Hong, Tao and Pinson, Pierre and Fan, Shu and Zareipour, Hamidreza and Troccoli, Alberto and Hyndman, Rob J.},
  title   = {Probabilistic energy forecasting: {Global} {Energy} {Forecasting} {Competition} 2014 and beyond},
  journal = {International Journal of Forecasting},
  year    = {2016},
  volume  = {32},
  number  = {3},
  pages   = {896--913},
  doi     = {10.1016/j.ijforecast.2016.02.001}
}

@article{haben2021gb,
  author  = {Haben, Stephen and Caudron, Julien and Verma, Jake},
  title   = {Probabilistic day-ahead wholesale price forecast: A case study in {Great} {Britain}},
  journal = {Forecasting},
  year    = {2021},
  volume  = {3},
  number  = {3},
  pages   = {596--632},
  doi     = {10.3390/forecast3030038}
}

@article{osone2025ukqr,
  author  = {Osone, Yuki and Kodaira, Daisuke},
  title   = {Quantile regression for probabilistic electricity price forecasting in the {U.K.} electricity market},
  journal = {IEEE Access},
  year    = {2025},
  volume  = {13},
  pages   = {10083--10093},
  doi     = {10.1109/ACCESS.2025.3528450}
}

@article{krishnamurthy2018storage,
  author  = {Krishnamurthy, Dheepak and Uckun, Canan and Zhou, Zhi and Thimmapuram, Prakash R. and Botterud, Audun},
  title   = {Energy storage arbitrage under day-ahead and real-time price uncertainty},
  journal = {IEEE Transactions on Power Systems},
  year    = {2018},
  volume  = {33},
  number  = {1},
  pages   = {84--93},
  doi     = {10.1109/TPWRS.2017.2685347}
}

@article{oconnor2025review,
  author  = {O'Connor, Ciaran and Bahloul, Mohamed and Prestwich, Steven and Visentin, Andrea},
  title   = {The Evolution of Probabilistic Price Forecasting Techniques: A Review of the Day-Ahead, Intra-Day, and Balancing Markets},
  journal = {Energies},
  year    = {2026},
  volume  = {19},
  number  = {8},
  pages   = {1929},
  doi     = {10.3390/en19081929}
}

@inproceedings{romano2019cqr,
  url     = {https://papers.nips.cc/paper/2019/hash/5103c3584b063c431bd1268e9b5e76fb-Abstract.html},
  accessed = {10 July 2026},
  author    = {Romano, Yaniv and Patterson, Evan and Cand{\`e}s, Emmanuel J.},
  title     = {Conformalized quantile regression},
  booktitle = {Advances in Neural Information Processing Systems (NeurIPS)},
  year      = {2019},
  volume    = {32},
  pages     = {3543--3553}
}

@article{kath2021conformal,
  author  = {Kath, Christopher and Ziel, Florian},
  title   = {Conformal prediction interval estimation and applications to day-ahead and intraday power markets},
  journal = {International Journal of Forecasting},
  year    = {2021},
  volume  = {37},
  number  = {2},
  pages   = {777--799},
  doi     = {10.1016/j.ijforecast.2020.09.006}
}

@article{meinshausen2006qrf,
  url     = {https://www.jmlr.org/papers/v7/meinshausen06a.html},
  accessed = {10 July 2026},
  author  = {Meinshausen, Nicolai},
  title   = {Quantile Regression Forests},
  journal = {Journal of Machine Learning Research},
  year    = {2006},
  volume  = {7},
  pages   = {983--999}
}

@article{brusaferri2025conformal,
  author  = {Brusaferri, Alessandro and Ballarino, Andrea and Grossi, Luigi and Laurini, Fabrizio},
  title   = {On-line conformalized neural networks ensembles for probabilistic forecasting of day-ahead electricity prices},
  journal = {Applied Energy},
  year    = {2025},
  volume  = {398},
  pages   = {126412},
  doi     = {10.1016/j.apenergy.2025.126412}
}

@article{oconnor2025conformal,
  author  = {O'Connor, Ciaran and Bahloul, Mohamed and Rossi, Roberto and Prestwich, Steven and Visentin, Andrea},
  title   = {Conformal prediction for electricity price forecasting in the day-ahead and real-time balancing market},
  journal = {Energy and AI},
  year    = {2025},
  volume  = {21},
  pages   = {100571},
  doi     = {10.1016/j.egyai.2025.100571}
}

@article{maciejowska2024hybridqra,
  author  = {Maciejowska, Katarzyna and Serafin, Tomasz and Uniejewski, Bartosz},
  title   = {Probabilistic forecasting with a hybrid {Factor-QRA} approach: Application to electricity trading},
  journal = {Electric Power Systems Research},
  year    = {2024},
  volume  = {234},
  pages   = {110541},
  doi     = {10.1016/j.epsr.2024.110541}
}

@article{lipiecki2024postprocessing,
  author  = {Lipiecki, Arkadiusz and Uniejewski, Bartosz and Weron, Rafa{\l}},
  title   = {Postprocessing of point predictions for probabilistic forecasting of day-ahead electricity prices: The benefits of using Isotonic Distributional Regression},
  journal = {Energy Economics},
  year    = {2024},
  volume  = {139},
  pages   = {107934},
  doi     = {10.1016/j.eneco.2024.107934}
}

@article{hagfors2016uk,
  author  = {Hagfors, Lars Ivar and Bunn, Derek and Kristoffersen, Eline and Staver, Tiril Toftdahl and Westgaard, Sjur},
  title   = {Modeling the {UK} electricity price distributions using quantile regression},
  journal = {Energy},
  year    = {2016},
  volume  = {102},
  pages   = {231--243},
  doi     = {10.1016/j.energy.2016.02.025}
}

@misc{desnz2025energytrends,
  author       = {{Department for Energy Security and Net Zero}},
  title        = {Digest of {United Kingdom} Energy Statistics 2025},
  year         = {2025},
  howpublished = {UK Government Official Statistics},
  url          = {https://assets.publishing.service.gov.uk/media/688b52e67b28774ca0437f09/DUKES_2025_Chapters_1-7.pdf},
  accessed = {10 July 2026},
}

@misc{epexspot2026basics,
  author       = {{EPEX SPOT}},
  title        = {Basics of the Power Market},
  year         = {2026},
  howpublished = {EPEX SPOT market information},
  url          = {https://www.epexspot.com/en/basicspowermarket},
  accessed = {28 June 2026},
}

@misc{epexspot2026annualresults,
  author       = {{EPEX SPOT}},
  title        = {{EPEX SPOT} Annual Power Trading Results 2025},
  year         = {2026},
  howpublished = {Annual trading results},
  url          = {https://www.epexspot.com/sites/default/files/download_center_files/2026-01-19_EPEX%20SPOT_Annual%20Power%20Trading%20Results%202025_final_0.pdf},
  accessed     = {21 July 2026},
}

@misc{epexspot2026marketresults,
  author       = {{EPEX SPOT}},
  title        = {Market Results: {GB DAA 1 (60')} Day-Ahead Auction},
  year         = {2026},
  howpublished = {EPEX SPOT market data},
  url          = {https://www.epexspot.com/en/market-results},
  accessed     = {21 July 2026},
}

@misc{epexspot2026tradingproducts,
  author       = {{EPEX SPOT}},
  title        = {Trading Products},
  year         = {2026},
  howpublished = {EPEX SPOT market information},
  url          = {https://www.epexspot.com/en/tradingproducts},
  accessed = {28 June 2026},
}

@misc{elexon2026insights,
  author       = {{Elexon}},
  title        = {Insights Solution API},
  year         = {2026},
  howpublished = {Balancing and Settlement Code Company data service},
  url          = {https://bmrs.elexon.co.uk/api-documentation/introduction},
  accessed     = {9 August 2026}
}

@misc{neso2026dataportal,
  author       = {{National Energy System Operator}},
  title        = {Data Portal},
  year         = {2026},
  howpublished = {Electricity-system data service},
  url          = {https://www.neso.energy/data-portal},
  accessed     = {9 August 2026}
}

@misc{neso2026opmr,
  author       = {{National Energy System Operator}},
  title        = {Daily Operational Planning Margin Requirement},
  year         = {2026},
  howpublished = {NESO Data Portal dataset},
  url          = {https://www.neso.energy/data-portal/daily-opmr},
  accessed     = {9 August 2026}
}

@misc{investing2026ukgas,
  author       = {{Investing.com}},
  title        = {{UK} Natural Gas Futures Historical Data},
  year         = {2026},
  howpublished = {Historical futures-price download},
  url          = {https://uk.investing.com/commodities/natural-gas-historical-data?cid=1057002},
  accessed     = {24 July 2026}
}

@misc{investing2026ukcarbon,
  author       = {{Investing.com}},
  title        = {{UK} Emissions Allowances Energy c1 Futures Historical Data},
  year         = {2026},
  howpublished = {Historical futures-price download},
  url          = {https://uk.investing.com/commodities/uk-emissions-allowances-energy-c1-futures-historical-data},
  accessed     = {24 July 2026}
}

@inproceedings{trebbien2023lstm,
  author    = {Trebbien, Julius and P{\"u}tz, Sebastian and Sch{\"a}fer, Benjamin and Nyg{\aa}rd, Heidi S. and Rydin Gorj{\~a}o, Leonardo and Witthaut, Dirk},
  title     = {Probabilistic Forecasting of Day-Ahead Electricity Prices and their Volatility with {LSTMs}},
  booktitle = {2023 IEEE PES Innovative Smart Grid Technologies Europe (ISGT EUROPE)},
  year      = {2023},
  address   = {Grenoble, France},
  pages     = {1--5},
  publisher = {IEEE},
  doi       = {10.1109/ISGTEUROPE56780.2023.10407112}
}

@inproceedings{gibbs2021aci,
  url     = {https://proceedings.neurips.cc/paper/2021/hash/0d441de75945e5acbc865406fc9a2559-Abstract.html},
  accessed = {10 July 2026},
  author    = {Gibbs, Isaac and Cand{\`e}s, Emmanuel J.},
  title     = {Adaptive Conformal Inference Under Distribution Shift},
  booktitle = {Advances in Neural Information Processing Systems (NeurIPS)},
  year      = {2021},
  volume    = {34},
  pages     = {1660--1672}
}

@article{barber2023nexcp,
  author  = {Barber, Rina Foygel and Cand{\`e}s, Emmanuel J. and Ramdas, Aaditya and Tibshirani, Ryan J.},
  title   = {Conformal prediction beyond exchangeability},
  journal = {The Annals of Statistics},
  year    = {2023},
  volume  = {51},
  number  = {2},
  pages   = {816--845},
  doi     = {10.1214/23-AOS2276}
}

@article{liu2023lstnetquantile,
  author  = {Liu, Huixin and Shen, Xiaodong and Tang, Xisheng and Liu, Junyong},
  title   = {Day-Ahead Electricity Price Probabilistic Forecasting Based on {SHAP} Feature Selection and {LSTNet} Quantile Regression},
  journal = {Energies},
  year    = {2023},
  volume  = {16},
  number  = {13},
  pages   = {5152},
  doi     = {10.3390/en16135152}
}

@inproceedings{angelopoulos2023conformalpid,
  url     = {https://proceedings.neurips.cc/paper_files/paper/2023/hash/47f2fad8c1111d07f83c91be7870f8db-Abstract-Conference.html},
  accessed = {10 July 2026},
  author    = {Angelopoulos, Anastasios N. and Cand{\`e}s, Emmanuel J. and Tibshirani, Ryan J.},
  title     = {Conformal {PID} Control for Time Series Prediction},
  booktitle = {Advances in Neural Information Processing Systems},
  year      = {2023},
  volume    = {36},
  pages     = {23047--23074},
  doi       = {10.52202/075280-1000}
}

@article{jensen2024encqr,
  author  = {Jensen, Viktor and Bianchi, Filippo Maria and Anfinsen, Stian Normann},
  title   = {Ensemble Conformalized Quantile Regression for Probabilistic Time Series Forecasting},
  journal = {IEEE Transactions on Neural Networks and Learning Systems},
  year    = {2024},
  volume  = {35},
  number  = {7},
  pages   = {9014--9025},
  doi     = {10.1109/TNNLS.2022.3217694}
}

@article{chen2025oanqr,
  author  = {Chen, Zhengganzhe and Zhang, Bin and Du, Chenglong and Yang, Chunhua and Gui, Weihua},
  title   = {Outlier-adaptive-based non-crossing quantiles method for day-ahead electricity price forecasting},
  journal = {Applied Energy},
  year    = {2025},
  volume  = {382},
  pages   = {125328},
  doi     = {10.1016/j.apenergy.2025.125328}
}

@article{benamor2025rnnlinear,
  author  = {Ben Amor, Souhir and Ziel, Florian},
  title   = {Recurrent neural networks with linear structures for electricity price forecasting},
  journal = {Renewable and Sustainable Energy Reviews},
  year    = {2026},
  volume  = {231},
  pages   = {116773},
  doi     = {10.1016/j.rser.2026.116773}
}

@article{koenker1978regression,
  author  = {Koenker, Roger and Bassett, Gilbert},
  title   = {Regression Quantiles},
  journal = {Econometrica},
  year    = {1978},
  volume  = {46},
  number  = {1},
  pages   = {33--50},
  doi     = {10.2307/1913643}
}

@book{vovk2005algorithmic,
  author    = {Vovk, Vladimir and Gammerman, Alexander and Shafer, Glenn},
  title     = {Algorithmic Learning in a Random World},
  publisher = {Springer},
  address   = {New York},
  year      = {2005},
  doi       = {10.1007/b106715}
}

@article{hochreiter1997lstm,
  author  = {Hochreiter, Sepp and Schmidhuber, J{\"u}rgen},
  title   = {Long Short-Term Memory},
  journal = {Neural Computation},
  year    = {1997},
  volume  = {9},
  number  = {8},
  pages   = {1735--1780},
  doi     = {10.1162/neco.1997.9.8.1735}
}

@article{elman1990structure,
  author  = {Elman, Jeffrey L.},
  title   = {Finding Structure in Time},
  journal = {Cognitive Science},
  year    = {1990},
  volume  = {14},
  number  = {2},
  pages   = {179--211},
  doi     = {10.1207/s15516709cog1402_1}
}

@article{christoffersen1998interval,
  author  = {Christoffersen, Peter F.},
  title   = {Evaluating Interval Forecasts},
  journal = {International Economic Review},
  year    = {1998},
  volume  = {39},
  number  = {4},
  pages   = {841--862},
  doi     = {10.2307/2527341}
}

@misc{ofgem2026wholesale,
  author       = {{Ofgem}},
  title        = {Wholesale energy costs and your bills},
  year         = {2026},
  howpublished = {Ofgem},
  url          = {https://www.ofgem.gov.uk/blog/wholesale-energy-costs-and-your-bills},
  accessed = {30 June 2026},
}

@article{kean2026wind,
  author  = {Kean, James and O'Sullivan, Aidan},
  title   = {Forecasting onshore wind generation in {European} bidding zones using deep learning},
  journal = {npj Clean Energy},
  year    = {2026},
  volume  = {2},
  pages   = {19},
  doi     = {10.1038/s44406-026-00036-6},
}

@article{zakeri2023gas,
  author  = {Zakeri, Behnam and Staffell, Iain and Dodds, Paul E. and Grubb, Michael and Ekins, Paul and J{\"a}{\"a}skel{\"a}inen, Jaakko and Cross, Samuel and Helin, Kristo and Castagneto-Gissey, Giorgio},
  title   = {The role of natural gas in setting electricity prices in {E}urope},
  journal = {Energy Reports},
  year    = {2023},
  volume  = {10},
  pages   = {2778--2792},
  doi     = {10.1016/j.egyr.2023.09.069}
}

@inproceedings{wen2017mqrnn,
  author    = {Wen, Ruofeng and Torkkola, Kari and Narayanaswamy, Balakrishnan and Madeka, Dhruv},
  title     = {A Multi-Horizon Quantile Recurrent Forecaster},
  booktitle = {31st Conference on Neural Information Processing Systems (NIPS 2017), Time Series Workshop},
  year      = {2017},
  address   = {Long Beach, CA, USA},
  doi       = {10.48550/arXiv.1711.11053}
}

@article{bottieau2020imbalance,
  author  = {Bottieau, J{\'e}r{\'e}mie and Hubert, Louis and De Gr{\`e}ve, Zacharie and Vall{\'e}e, Fran{\c{c}}ois and Toubeau, Jean-Fran{\c{c}}ois},
  title   = {Very-Short-Term Probabilistic Forecasting for a Risk-Aware Participation in the Single Price Imbalance Settlement},
  journal = {IEEE Transactions on Power Systems},
  year    = {2020},
  volume  = {35},
  number  = {2},
  pages   = {1218--1230},
  doi     = {10.1109/TPWRS.2019.2940756},
}

@inproceedings{kingma2015adam,
  author    = {Kingma, Diederik P. and Ba, Jimmy},
  title     = {Adam: A Method for Stochastic Optimization},
  booktitle = {International Conference on Learning Representations},
  year      = {2015},
  url       = {https://arxiv.org/abs/1412.6980},
  accessed  = {25 August 2026}
}

@article{srivastava2014dropout,
  author  = {Srivastava, Nitish and Hinton, Geoffrey and Krizhevsky, Alex and
             Sutskever, Ilya and Salakhutdinov, Ruslan},
  title   = {Dropout: A Simple Way to Prevent Neural Networks from Overfitting},
  journal = {Journal of Machine Learning Research},
  volume  = {15},
  pages   = {1929--1958},
  year    = {2014},
  url     = {https://jmlr.org/papers/v15/srivastava14a.html},
  accessed = {25 August 2026}
}

\end{document}